\documentclass[aps,amsmath,amssymb,pra,twocolumn,footinbib,superscriptaddress,floatfix]{revtex4-2}

\usepackage{graphicx}% Include figure files
\usepackage{dcolumn}% Align table columns on decimal point
\usepackage{bm}% bold math
\usepackage{footmisc}
\usepackage{mathtools}
\usepackage{framed}
\usepackage{mathrsfs}
\usepackage{lipsum}
\usepackage{makecell}
\usepackage{physics}

\usepackage{hyperref}
\hypersetup{%
colorlinks=true,
linkcolor=black,
citecolor=NavyBlue,
urlcolor=NavyBlue,
plainpages=false
}
\usepackage[dvipsnames]{color,xcolor,colortbl}

\newcommand{\N}{\mathbb{N}}

\newcommand{\R}{\mathbb{R}}
\newcommand{\C}{\mathbb{C}}

\def\d{{\rm d}}

\def\>{\rangle}
\def\<{\langle}

\newcommand{\bs}[1]{\boldsymbol{#1}}
\newcommand{\map}[1]{\mathcal{#1}}
\newtheorem{theo}{Theorem}

\newtheorem{cor}{Corollary}

\begin{document}

\preprint{APS/123-QED}

\title{Quantum illumination networks}

\author{Xiaobin Zhao}
\email{xzhao721@usc.edu}
\affiliation{Ming Hsieh Department of Electrical and Computer Engineering,
University of Southern California, Los Angeles, CA 90089, USA}

\author{Zheshen Zhang}
\affiliation{
Department of Electrical Engineering and Computer Science,
University of Michigan, Ann Arbor, MI 48109, USA
}

\author{Quntao Zhuang}
\email{qzhuang@usc.edu}
\affiliation{Ming Hsieh Department of Electrical and Computer Engineering,
University of Southern California, Los Angeles, CA 90089, USA}
\affiliation{Department of Physics and Astronomy, University of Southern California, Los Angeles, CA 90089, USA}

%communications physics, deadline Sep 4 2024

\nopagebreak

\begin{abstract}
%Quantum illumination is an entanglement-based target detection protocol that provides quantum advantages despite the presence of entanglement-breaking noise. However, the advantage of traditional quantum illumination protocols relies on an extremely low total transmitted power and high background noise, substantially limiting its practical relevance.  In this work, we propose a quantum illumination network with a transmitter array and a single receiver antenna that offers a significant advantage in multi-parameter estimation of targets, even in low noise and high signal power regions. Due to network interference and the multi-parameter nature of the problem, entanglement's advantage applies to zero background noise and high brightness region, applicable to lidar scenarios. Due to multiple transmitters, the total power can be as high as milliWatt, relevant to practical systems. More importantly, when the number of probes is limited, the quantum advantage is huge as classical protocols fail to unambiguously estimate all parameters. Overall, a quantum illumination network provides a path for quantum sensing advantages at practically relevant power and noise levels, and to much wider frequency regions including both radar and lidar.

Quantum illumination is an entanglement-based target detection protocol that provides quantum advantages despite the presence of entanglement-breaking noise. However, the advantage of traditional quantum illumination protocols is limited to impractical scenarios with low transmitted power and simple target configurations. In this work, we propose a quantum illumination network to overcome the limitations, via designing a transmitter array and a single receiver antenna. Thanks to multiple transmitters, quantum advantage is achieved even with a high total transmitted power. Moreover, for single-parameter estimation, the advantage of network over a single transmitter case increases with the number of transmitters before saturation. At the same time, complex target configurations with multiple unknown transmissivity or phase parameters can be resolved. Despite the interference of different returning signals at the single antenna and photon-loss due to multiple-access channel, we provide two types of measurement design, one based on parametric-amplification and one based on the correlation-to-displacement conversion (CtoD) to achieve a quantum advantage in estimating all unknown parameters. We also generalize the parameter estimation scenario to a general hypothesis testing scenario, where the six-decibel quantum illumination advantage is achieved at a much greater total probing power.

\end{abstract}

\maketitle

\section{Introduction}

Entanglement is a unique feature of quantum physics that brings benefits in information processing tasks~\cite{zhang2024entanglement}.
Quantum illumination (QI) is an example where entanglement provides advantages in sensing tasks even when it is eventually destroyed by noise and loss during the sensing process~\cite{lloyd2008enhanced,tan2008quantum}. The protocol sends out a signal to probe the target while storing the entangled idler for reference. Upon return of the noisy signal, a measurement is performed on both the return and the idler to determine the properties of the target.

QI came as a conceptual surprise, and much effort has been devoted towards making the quantum advantage practically relevant. To begin with, the original paper by Tan {\em et al.}~\cite{tan2008quantum} showed quantum advantage with performance bounds and left the measurement design problem open. The initial design based on off-the-shelf components of the parametric amplifier provides sub-optimal quantum advantages~\cite{guha2009gaussian}, which has been demonstrated in the optical domain~\cite{zhang2015entanglement,hao2022demonstration} and more recently in the microwave domain~\cite{assouly2023quantum}. The optimal measurement has been subsequently proposed, with the sum-frequency generation process~\cite{zhuang2017optimum}  and via the \emph{correlation-to-displacement} (CtoD) conversion~\cite{shi2022fulfilling,shi2024optimal}. With the development of the CtoD concept, simplified sub-optimal receivers based on heterodyne-homodyne is proposed to bring hope to practical microwave implementations~\cite{reichert2023quantum}.

While the challenges in the measurement design have been relaxed, many other issues plague the practical relevance of QI, as summarized in Refs.~\cite{shapiro2020quantum,karsa2023quantum}. One problem regards the fact that the original protocol only detects the presence and absence of a single target in a spatiotemporal bin, which is far away from real radar detection scenarios. Efforts in extending the applicability of QI have shown that advantage in ranging can be possible~\cite{zhuang2021quantum,zhuang2022ultimate}, where the target can be at different locations along a single direction. However, the more important problem is that QI is considering an energy constraint that is far from realistic. As the brightness---the photon flux per bandwidth---needs to be less than unity to enable quantum advantage, the power of the transmitter is extremely low given the gigahertz bandwidth available at microwave frequency.

\begin{figure}[]
\includegraphics[width=0.53\textwidth,trim=40 2 2 2,clip]{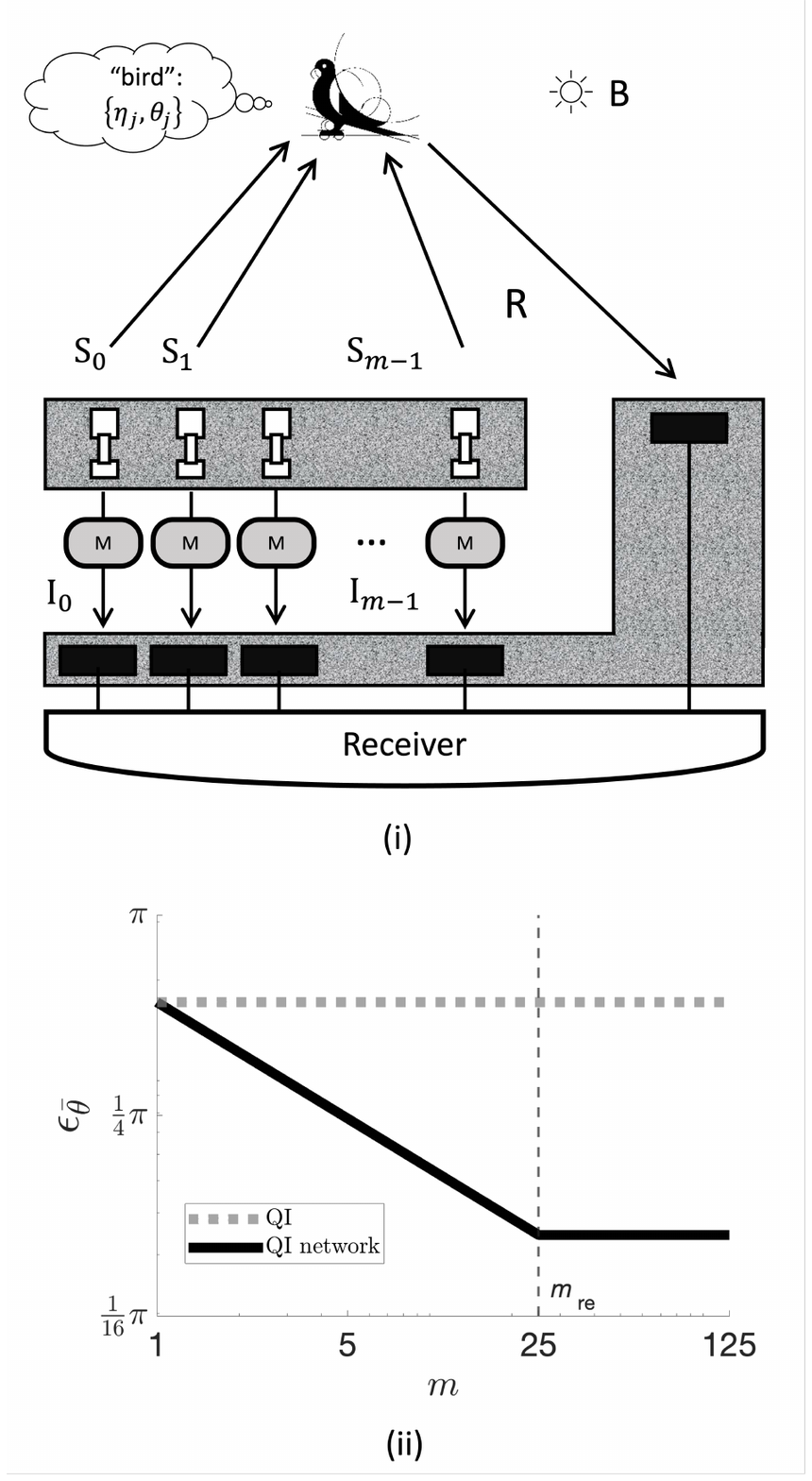}
\caption[]{(i)Quantum illumination network set-up. A transmitter array sends out multiple probes $\{S_j\}$ to multiple targets (or different aspects of the same target) and then stores the corresponding entangled idlers $\{I_j\}$ in quantum memory marked by $M$. The physical parameters that need to be identified correspond to the phase $\{\theta_j\}$ and reflectivity $\{\eta_j\}$ imprinted on the return state. With a single antenna to receive the returning mode $R$, the experimenter can conduct multi-parameter quantum estimation and hypothesis testing under the influence of background noise (marked by $B$). (ii) A simple depiction of the discrepancy in estimation performance between a QI network and a traditional QI protocols. Here we illustrate the root-mean-square error in estimating an average of multiple phases with reflectivity ratio $\eta\sim 0.5$, the photon numbers $N_{\rm S}=0.5$ and $N_{\rm B}=32$. 
}\label{fig:configuration}
\end{figure}

In this work, we propose a quantum illumination network to resolve the above low-power and single-target constraints of the original QI protocols. While a single transceiver can only detect a single target, in the QI network, multiple targets are simultaneously probed with a network of transmitters, and the return is detected by a single receiver antenna (see Fig.~\ref{fig:configuration}i). Due to multiple transmitters probing the same region, the total probing power can be large. As a result, for single target case, the advantage of QI network over the original single-transmitter QI protocol increases with the number of transmitters before saturation (see Fig.~\ref{fig:configuration}ii). The inevitable interference of different returning signals creates a challenge to the reception end. Surprisingly, we show that even with a single receiver, one is able to overcome the interference problem and achieve estimation precision advantages over the best classical strategies in the estimation of multiple phases or transmissivities. We also provide the measurement strategy to achieve quantum advantage based on parametric amplification~\cite{guha2009gaussian} or the CtoD conversion~\cite{shi2022fulfilling,shi2024optimal}. 
\\[-8pt] 

%\QZ{four major findings thanks to a network: (1)much higher total power due to parallel transmitters. also relaxes brightness constraints, high brightness can also be advantageous. (2)extend QI to network scenario}

The remainder of the paper is organized as follows. In Section~\ref{sec:QI_SET_UP}, we describe the framework of the QI network. Section~\ref{sec:application} focuses on the practical relevance of our proposal. Then, we examine the classical benchmark of the QI network in Section~\ref{sec:classical_benchmark} and introduce the measurement protocols in Section~\ref{sec:mesurement_design}. The sensing superiority of the QI network is demonstrated in the analysis of multiple-phase sensing in Section~\ref{sec:phase_sensing} and in the classification of reflectivity patterns in Section~\ref{sec:pattern}. Following that, we discuss a potential advantage of non-asymptotic parameter estimation in Section~\ref{sec:potential_opt}.   Finally, the conclusions are drawn in Section~\ref{sec:conclusions}.

\section{Quantum illumination network set-up} 
\label{sec:QI_SET_UP}

A shown in Figure \ref{fig:configuration}, a quantum illumination network consists of an array of $m$ transmitters, each emits a signal-idler pair $\{{\rm S}_j,{\rm I}_j\}$ in a two-mode squeezed vacuum (TMSV) state \cite{weedbrook2012gaussian} 
\begin{align}
|\Phi\>=\exp\left[\frac 1 2\,  \text{arcsinh}\sqrt{N_{\rm S}}\left(a_{{\rm S}_j} a_{{\rm I}_j}-a_{{\rm S}_j}^\dag a_{{\rm I}_j}^\dag \right)\right]|0\>_{{\rm S}_j}|0\>_{{\rm I}_j},
\end{align}
where $N_{\rm S}$ is the average photon number of $S_j$ or $I_j$, $a_{{\rm S}_j}(a_{{\rm I}_j})$ and $a_{{\rm S}_j}^\dag(a_{{\rm I}_j}^\dag)$ denote the annihilation and creation operators of the signal (idler) mode, respectively, and $|0\>$ is the vacuum state. Then, the experimenter stores the idlers, $\{a_{{\rm I}_j}\}_{j=0}^{m-1}$, and sends the $m$ signal modes, $\{a_{{\rm S}_j}\}_{j=0}^{m-1}$, to the target. The experimental realization of the transmitter array can be potentially achieved by nano-antenna array~\cite{llatser2012graphene,kavitha2022graphene}, in particular at the higher frequency end of radar detection.

At the common receiver, a single return mode is received, with interference between all return probes. The $m$ transmitters are able to excite more spatial modes of the target, which may have non-zero overlap with the receiver spatial mode. As a simplified model, we assume the maximum number of spatial modes of the target (or multiple targets) being excited is $m_{\rm re}$ and different transmitters excite different spatial modes. Therefore, the return mode is given by the input-output relation
\begin{align}\label{eq2}
a_{\rm R}=\sum_{j=0}^{m_{\rm re}-1} \frac{1}{\sqrt{m_{\rm re}}}a_{{\rm R}_j},
\end{align}
which forms a multiple-access channel~\cite{winter2001capacity,shi2021entanglement} from $m_{\rm re}$ senders and a single receiver. Note that we assume that the other $m_{\rm re}-1$ output modes other than $\hat{a}_R$ are not accessible to the receiver end. When there are $m\le m_{\rm re}$ transmitters, the rest $m_{\rm re}-m$ transmitter modes are in vacuum.
% To ease the description, we can express the multiple-access channel as a two-step process. First, the interference is described by a beamsplitter transform,
% \begin{align}
% a_{\rm R}=\sum_j \sqrt{w_j}\, a_{{\rm R}_j},
% \end{align}
% where $\{w_j\}$ are known complex numbers and normalized $\sum_{j=1}^m |w_j|=1$. \XZ{Although it is not possible to concentrate the energy of the second moment with passive operations (or basis rotation), the $m$-to-$1$ channel of target reflection can be modeled as a projection as in discussions of the cloning channel \cite{chiribella2014optimal}. On this account, the constraint on the weights $\{\omega_j\}$ may be removed to allow them to be mutually independent and unnormalized. Then, they  function equivalently to $\{\eta_j\}$ while satisfying the requirement $\sum_{j=1}^m |w_j|=m\,\overline \omega$ with $1/m<\overline \omega\le 1$. } 
Here we define each virtual individual return $a_{{\rm R}_j}$ to model the loss and noise in the channel~\cite{yuen1976two,tan2008quantum,yuen2009classicalization},
\begin{align}\label{thermal-loss}
a_{{\rm R}_j}=e^{i\theta_j}\sqrt{\eta_j} \,{a}_{{\rm S}_j}+\sqrt{1-\eta_j}\,a_{{\rm B}_j}, \, 1\le j \le m,
\end{align} 
where $\eta_j\in [0,1]$ is the reflectivity, $ \theta_j\in [0,2\pi)$ is the phase angle and the noise mode $a_{{\rm B}_j}$ has an average photon number $N_{\rm B}/(1-\eta_j)\gg 1$. Typically, the actual number of transmitters $m \ll m_{\rm re}$, therefore we focus on the case where $m<m_{\rm re}$. Here, we adopt the premise of reflectivity-independent noise $N_{\rm B}$ as in the original approach~\cite{tan2008quantum}, which may arise from realistic scenarios such as low reflectivity~\cite{tan2008quantum}, bright thermal-noise bath~\cite{tan2008quantum,wilde2017gaussian}, Gaussian measurement~\cite{karsa2020noisy}, and Gaussian additive noise channels~\cite{weedbrook2012gaussian}.

Here the targets are described by $m$ phase shifts $\{\theta_j\}_{j=0}^{m-1}$ and $m$ transmissivities $\{\eta_j\}_{j=0}^{m-1}$. These parameters can in general model different targets in the same region, or different parts of a single target. As a special case, the $m$ sets of parameters can also be equal, representing a degenerate case where only a single target is being considered. Such a degenerate case represents a spatial multiplexing at the transmitter. Due to multiple transmitters, the total transmitted average photon number is increased to $mN_{\rm S}$; at the same time, as we assume a single receiver, the received photon number 
\begin{align} \label{eq4}
\expval{a_{\rm R}^\dagger a_{\rm R}}=\sum_{j=0}^{m-1} \frac{1}{m_{\rm re}}(\eta_j N_{\rm S}+N_{\rm B})
\end{align} 
increases linearly with $m$ before saturation at large $m=m_{\rm re}$ transmitters. As a result, the performance of the QI network increases with the number of transmitters $m$ before saturation at $m=m_{\rm re}$. As exemplified in Fig.~\ref{fig:configuration}ii, the root-mean-square estimation error in single parameter estimation of QI network improves as the number of transmitters increase. The performance evaluation utilizes Eq.~\eqref{wrmse} in the degenerate case, as we detail in Section~\ref{sec:phase_sensing}.

The final measurement is applied on the return mode R jointly with $m$ idler modes, $\{{\rm I}_j\}_{j=0}^{m-1}$.
The resulting $(m+1)$-mode is in a zero-mean Gaussian state with quadrature covariance matrix (see basic definitions with natural units $\hbar =2$ in Ref. \cite{weedbrook2012gaussian}):
\begin{align}\label{CoV}
\bs V=\left(\begin{matrix}
\left(2N_{\rm B}'+1\right)\mathbb I_2 &S_{0}^{\rm T}&\cdots &S_{m-1}^{\rm T}\\
S_{0} & (2N_{\rm S}+1)\mathbb I_2 & \cdots &0\\
\vdots  &\vdots& \ddots   &\vdots \\
S_{m-1}& 0& \cdots  &(2N_{\rm S}+1)\mathbb I_2 \\
\end{matrix}\right),
\end{align}
where %N_{\rm B}':=\sum_{j=0}^{m-1} \omega_j(N_{\rm B}+ \eta_j N_{\rm S})$ 
$N_{\rm B}':= N_{\rm B}+ \sum_{j=0}^{m-1}\eta_j N_{\rm S}/m$ refers to the adjusted background photon number, $\mathbb I_\ell $ is the identity matrix of dimension $\ell$, %$\left\{S_{j}:= 2\,\text{Re}\left(\sqrt{\omega_{j}}\right)\sqrt{\eta_{j} N_{\rm S}(N_{\rm S}+1)}\mathbb Z\mathbb R_{j}^{\rm T}\right\}$
$\left\{S_{j}:= 2\,\sqrt{\eta_{j} N_{\rm S}(N_{\rm S}+1)/m_{\rm re}}\,\mathbb Z\mathbb R_{j}^{\rm T}\right\}$ are $2\times 2$ matrices defined by $\mathbb R_j =\cos \theta_j  \mathbb I_2 -i \sin \theta_j  \mathbb Y$ with $\mathbb Z$ and $\mathbb Y$ being the Pauli matrices.

To achieve the desired high precision in sensing, one sends out $\nu$ mode pairs, $\{{\rm S}_j^{(n)}, {\rm I}_j^{(n)}\}_{n=0}^{\nu-1}$, at each transmitter $0\le j \le m-1$ and therefore receive modes $R^{(n)}$'s. Such a repetition is typically realized by broadband probes, where $\nu=BT$ is the time-bandwidth product for bandwith $B$ and pulse duration $T$. We can also introduce the virtual received modes $R_j^{(n)}$, similar to the single pair case in Eq.~\eqref{thermal-loss}. The multiple mode pairs can come from the large time-bandwidth product of the spontaneous parametric down-conversion source that generates the TMSVs.
\\[-8pt]

Finally, the performance of sensing is characterized by the square root of the weighted mean-square error (rWMSE) 
\begin{equation}
\epsilon_{\bs \phi }:=\min_{\{M_a\}} \sqrt{ \sum_{j=0}^{m-1}\sum_a\Tr[\rho_{\bs \phi}M_a]\left[\hat{\phi}_j(a)-{ \phi}_j\right]^2/m}
\end{equation} 
where $\rho_{\bs \phi}$ refers to $\nu$ copies of output states from the QI process, the parameters of interest are $\bs \phi=(\theta_0,\cdots,\theta_{m-1})^T$ for multiple-phase sensing or $\bs \phi=(\eta_0,\cdots,\eta_{m-1})^T$ for reflectivity sensing, $\{M_a\}$ refer to the positive operator-valued measure (POVM) associated to $\nu$-rounds of measurement, the estimator $\{\hat{\phi}_j(a)\}$ are mapping from the measurement data $a$ (which could be multidimensional) to the parameters. 

Given the classical and quantum Crem\'er-Rao theorem \cite{fisher1925theory,rao1992information,cramer1999mathematical,helstrom1976quantum,holevo2011probabilistic,braunstein1994statistical,liu2020quantum}, the rWMSE of a measurement $\map M$ can be  bounded as follows: 
\begin{align}
\epsilon_{\map M,\bs\phi}\ge \sqrt{\frac 1 m \Tr \left[\map F_{\map M }^{-1}\right]}\ge \epsilon_{\bs\phi}\ge \sqrt{\frac 1 m \Tr \left[\map F^{-1}\right]},
\label{eq9s}
\end{align}
where $\map F_{\map M}$ is the classical Fisher information matrix based on the measurement results $p(a)=\Tr[\rho_{\bs \phi}M_a]$:
$\map F_{\map M,\phi_j,\phi_k}=\sum_a p(a)\left({\partial p(a)}/{\partial \phi_j}\right)\left({\partial p(a)}/{\partial \phi_k}\right),$ and $\map F$ is the quantum Fisher information matrix based on the output state $\rho_{\bs \phi}$: $\map F_{\phi_j,\phi_k}=\Tr[(L_iL_j+L_jL_i)\rho_{\bs \phi} ]/2$ with $L_i$ being the symmetric logarithmic derivative (SLD) defined by $\partial \rho_{\bs \phi}/\partial \phi_i=(\rho_{\bs \phi} L_i+L_i\rho_{\bs \phi})/2$.\\

The interference among multiple returning modes, as shown in Eq. (\ref{eq2}), will have distinct consequences in classical and quantum scenarios. For instance, in the trivial lossless and noiseless limit of $\eta_j=1,N_{\rm B}=0, \theta_j=0$, classical illumination (CI) with  coherent probe states ~\cite{serafini2017quantum,shapiro2020quantum} will lead to the output coherent state with an amplitude $m\sqrt{ N_{\rm S}/m_{\rm re}}$. In this scenario, the receiving photon number will have a scaling $\left\<a^\dag_{\rm R}a_{\rm R}\right\>_{\rm ci}\sim \mathcal O\left(m^2/m_{\rm re}\right)$. The photons of $m$ returning modes are regained in a single mode. In contrast, the receiving photon number in the QI network, given by Eq. (\ref{eq4}), has a scaling $\left\<a^\dag_{\rm R}a_{\rm R}\right\>_{\rm ci}\sim \mathcal O(m/m_{\rm re})$ with a much higher level of photon loss. Nevertheless, we will show that the QI network can still achieve quantum advantages in multi-parameter estimation and hypothesis testing.

\section{Application scenarios} 
\label{sec:application}

In contrast to previous QI protocols \cite{lloyd2008enhanced,tan2008quantum,shapiro2020quantum,karsa2023quantum}, the present approach exhibits significantly enhanced power of probing as a result of an $m$ tramsmitter system, hence increasing its practical applicability. Here we provide modeling for different sensing scenarios in terms of the parameter choices. We will obtain thermal noise from Bose-Einstein distribution $N_{\rm B}\propto 1/[\exp{(h f_p/k_B T)}-1]$, where $f_p$ is the center frequency, $T$ is the effective temperature of the system, $h$ is the Planck constant and $k_B$ is the Boltzmann constant. 
The total power of the QI network will be $P:=mN_{\rm S}B\,h\, f_p$, where $B$ refers to the bandwidth.

One parameter region of interest is the traditional microwave radar being considered in quantum illumination~\cite{tan2008quantum,zhuang2022ultimate,shapiro2020quantum}. We take W-band as an example, with $f_p=100$ GHz, bandwidth $B=10$ GHz and sky temperature $T=150$K, leading to thermal noise $N_{\rm B}\sim 32$. Another parameter region of interest is the THz-wave radar \cite{cooper2011thz,kokkoniemi2016discussion}. At THz frequency (wavelength $\sim 100$um), with the implementation of nano-antenna array~\cite{llatser2012graphene,kavitha2022graphene}, a large number of transmitters can be engineered, greatly enhancing the total power output over the single transmitter QI system~\cite{shapiro2020quantum}. 
%Then, if these transmitters emit an average photon number of $N_{\rm S}\sim 1$, operate at a carrier frequency of $f_p=6\,$ THz, and have a bandwidth of $B=0.5\,$ THz, the total power output will be about $2\,$ mW, many orders of magnitude higher than
%{\color{red}[see the comment at the outset. Can we fully justify the back-of-the-envelop estimation on the transmitted vs received power at the transmitter's and receiver's apertures? The diffraction at the terahertz bandwidth is large, rendering the highly directional connectivity between the transmitter and receiver implausible.]} \XZ{Without losing the generality, we assume the most favorable condition in classical scenario, where the displacements from $m$ probing modes are perfectly concentrated into one receiving mode through a passive Gaussian unitary process. } 
We consider $N_{\rm B}=0.6$ for 300K as THz radar is in short range. In this case, the QI network has the potential to exhibit advantages robust to the presence of various forms of sky noise~\cite{ippolito1981propagation,kokkoniemi2016discussion}.

On this ground, the QI networks can be applied to multi-parameter estimation and hypothesis testing, which feature circumstances when a macroscopic target cannot be effectively characterized by a single parameter due to its complexity. Moreover, the QI network may be taken into account when assessing the temporal evolution of the target. In the present work, the parameters of interest are phases and reflectivities imprinted on the returning state, each associated with the range or presence of the target. Finally, via the design of measurement protocols, these physical parameters and their statistical properties can be determined.

%\XZ{We may remove the following paragraph about lidar}

%Another application of the proposed system is at a lidar frequency of $300\,$THz, with a wavelength $\sim1000$nm. $N_{\rm B}\sim 0$. In this region, traditional quantum illumination has no advantage \cite{shapiro2020quantum}. Moreover, when it is lossy, quantum lidar has a very limited advantage due to the vacuum noise \cite{yuen2009classicalization,nair2011discriminating}. This is a region where typically no quantum advantage is possible. Here we show that in a multi-parameter estimation scenario, the quantum advantage is possible, as the output state of the QI network can not be replicated by finite implementations of classical protocols.
%\\[-0.5em]

%\QZ{from arXiv:2207.06609, we know the condition for advantage is $N_{\rm S}<N_{\rm B}/(1-\kappa)$, here maybe $N_{\rm S}$ also contributes to noise so $N_{\rm S}=0.8$ can have an advantage, so we see an advantage in different cases, even lidar}

\section{Classical benchmark}\label{sec:classical_benchmark}

As a benchmark for quantum advantage, we examine the minimal estimation error achievable by classical strategies, under the constraint that probes are prepared via a statistical mixture of coherent states~\cite{serafini2017quantum,shapiro2020quantum} with the same total signal power as the quantum case. We shall use the rWMSE to characterize the performance of estimation. In the degenerate case where all the phases have the same value, or where only the global information in the form of weighted average phase $\overline \theta'=\sum_j \sqrt{\eta_j/m_{\rm re}}\,\theta_j$ is unknown, we also adopt the root of the mean-square error (RMSE) directly to characterize the performance.

Based on this premise and the convexity of quantum Fisher information matrix (QFIM) \cite{takeoka2017fundamental,liu2020quantum,slaoui2022analytical}, we can obtain the following theorem.
{\theo\label{theo1}\textbf{(Asymptotic benchmark of phase sensing $(\nu\gg m)$)} In the classical illumination network, where the input is restricted to random mixtures of coherent states, its rWMSE of estimating $m$ independent phases satisfies the asymptotic bound 
\begin{align}\label{cla_rWMSE}
\epsilon_{{\bs\theta},\rm c}\ge 
%\begin{cases}
\sqrt{\frac{m_{\rm re}(2N_{\rm B}+1)}{4\nu N_{\rm S}\max_j \eta_j}},%&\nu\ge \frac m 2 %\\[0.6em]
%\sqrt{\frac{2\nu}{m} E_{{\rm 1-c}} +\left(1-\frac{2\nu}{m}\right) \frac{\pi^2}{3}},&\nu< \frac m 2 
%\end{cases}
\end{align}
}\\[0.5em]
A detailed proof of Theorem \ref{theo1} is illustrated in Appendix \ref{APP:benchmark_phase_sensing}. Note that it is impossible to achieve the equality of Ineq.~(\ref{cla_rWMSE}) by pure coherent states when the number of experiments is limited, particularly with the condition  $\nu < \lceil m/2\rceil$. This is caused by the fact that only mixed input states can output states with full-rank QFIMs. However, mixed input states such as amplified spontaneous emission (ASE) source are not optimal due to the convexity of QFIM (see detailed proof in Appendix \ref{APP:benchmark_phase_sensing}). Nevertheless, one can resolve this issue by conducting more than $m/2$ experiments, by which a full-rank QFIM can be achieved by a summation of non-full-rank QFIMs. We can prove that the rWMSE $\epsilon_{{\bs \theta},{\rm c}} =  \sqrt{m_{\rm re}(2N_{\rm B}+1)\sum_j \eta_j^{-1}/(4m\nu N_{\rm S})}$ is achievable with pure input states for the favorable condition $\nu= \ell m/2,\,$ for any positive integer $\ell$ and any positive even integer $m$. Specifically, the experimenter estimates two parameters $\theta_j$ and $\theta_k$ in each experiment, with an effective concentration of energy \cite{weedbrook2012gaussian,guha2011structured} and fine-tuned phases. Therefore, a diagonal QFIM $F_{\rm c}^{\{j,k\}}=2mN_{\rm S}/[m_{\rm re}(2N_{\rm B}+1)]\textbf{diag}(\eta_j,\eta_k)$ can be achieved~\cite{helstrom1969quantum,holevo2011probabilistic}. By repeating this experiment yet with different pairs of parameters, the experimenter is able to obtain the minimum rWMSE.

\section{Measurement design for quantum illumination}\label{sec:mesurement_design}

We propose two designs of measurement to achieve the quantum advantage of a QI network.

\begin{figure*}[t!]
\includegraphics[width=0.8\textwidth,trim=2 30 2 30,clip]{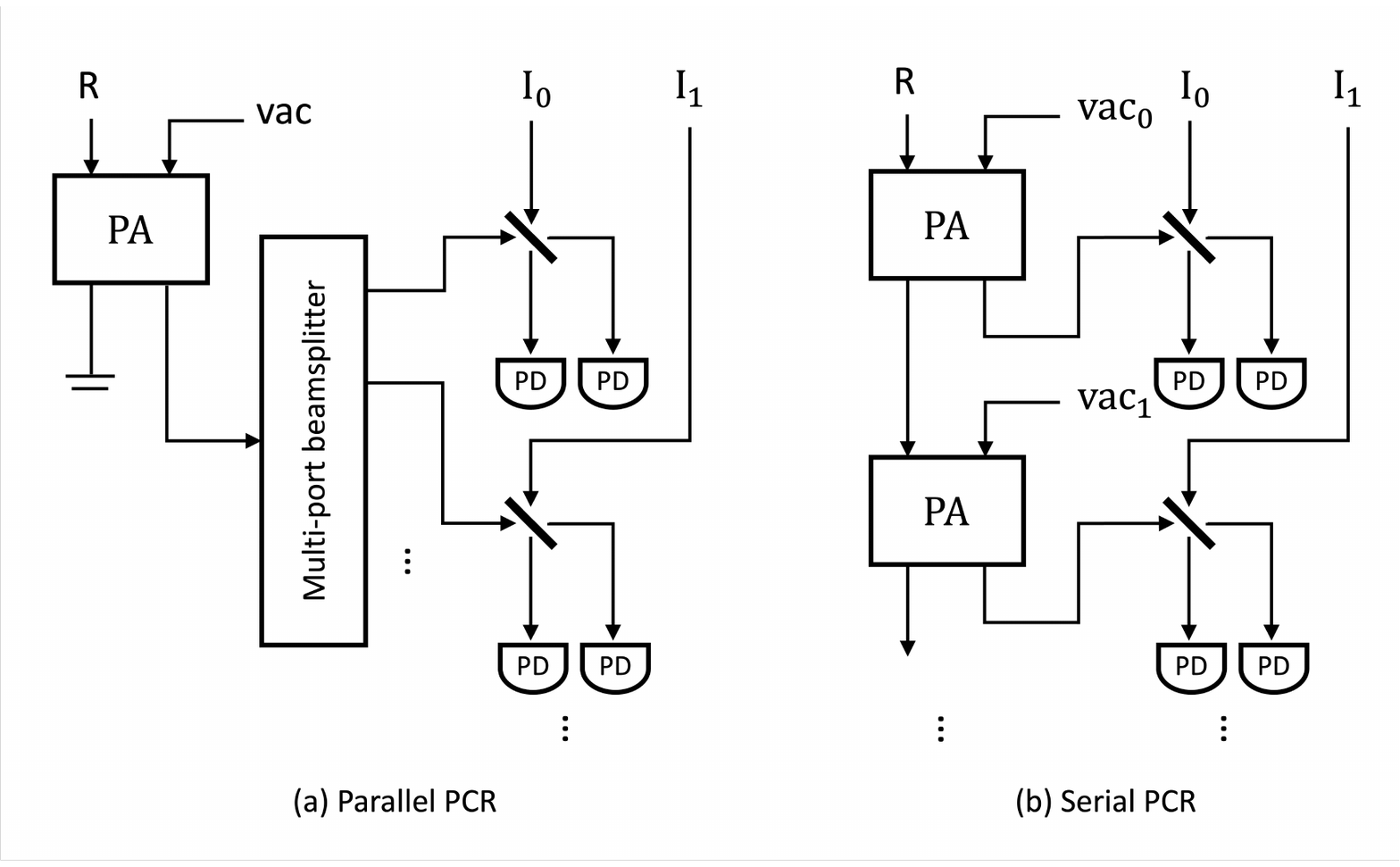}
\caption[]{{Schematic of two practical PA receiver designs. }(a) serial phase-conjugate receiver (sPCR), (b) parallel phase-conjugate receiver (pPCR). 
}\label{fig:OPA_PCR}
\end{figure*}

\subsection{Parametric amplifier network}\label{SEC:OPA_protocol}

The quantum illumination network allows the establishment of a multiple access channel \cite{winter2001capacity,shi2021entanglement}, where the follow-up measurement design of the $(m+1)$ modes can resort to the protocols based on parametric-amplifiers (PAs). For the sake of simplicity, we will examine two PA-based protocols: the parallel and serial phase-conjugate receiver (pPCR/ sPCR), where PA is adopted to perform a phase conjugation on the return. Both receivers provide a 3-dB quantum advantage in the error-probability exponent when it comes to discriminating between target's presence or absence~\cite{PhysRevA.80.052310}.

Specifically, the pPCR consists of the following steps (see schematic in Fig.~\ref{fig:OPA_PCR}(a) and detailed derivation in Appendix \ref{APP: QI-setup}): 
\begin{enumerate}
\item [(O1)] Conduct a joint PA operation on the return and vacuum to obtain the phase-conjugated return. 
%By discarding one output state, the covariance matrix is transformed as follows: \begin{align}\bs V=\left(\begin{matrix}(2\widetilde{N}_{{\rm B},g}+1)\mathbb I_2 &\sqrt {g-1} \mathbb Z S_{0}^{\rm T}&\cdots &\sqrt {g-1} \mathbb Z S_{m-1}^{\rm T}\\\sqrt {g-1} S_{0} \mathbb Z& (2N_{\rm S}+1)\mathbb I_2 & \cdots &0\\\vdots  &\vdots& \ddots   &\vdots \\\sqrt {g-1} S_{m-1}\mathbb Z& 0& \cdots  &(2N_{\rm S}+1)\mathbb I_2 \\\end{matrix}\right)\end{align}where $\widetilde N_{{\rm B},g}=(g-1)\left(N_{\rm B}'+1\right)$ is the effective photon number, $g\ge 1$ is the amplification gain. 
\item [(O2)] Distribute the phase conjugated return from the PA via a multi-port beamsplitter. 
\item [(O3)] Implement balanced beamsplitters that produce interference pairwisely on each portion of the phase conjugated return output from the multi-port beamsplitter and a set of idlers. 
\item [(O4)] Perform photodetection on the interfered modes and estimate physical parameters from the difference between total photon numbers from the two output ports of each balanced beamsplitter. 
\end{enumerate}

In addition to applying the pPCR protocol, the experimenter can implement multiple PA operations in a sequence. As illustrated in Figure \ref{fig:OPA_PCR} (b), the sPCR consists of the following steps (see details in Appendix \ref{APP: QI-setup}): 
\begin{enumerate}
\item [(O1$'$)] Same as O1.

\item [(O2$'$)] Consume one of the outputs from the PA operation to interact with one of the stored idler. Provide the other output and an additional vacuum to a subsequent PA. Repeat the aforementioned step for m
times in total.

\item [(O3$'$)] Implement balanced beamsplitter operations to pair-wisely generate interference between the PA outputs and the idlers. 

\item [(O4$'$)] Perform photodetection on the interfered modes and estimate the parameters of interest via detecting the photon count difference at each balanced beamsplitter. 
\end{enumerate}

Note that photon statistics from PA receivers are phase sensitive. To achieve the best performance, either prior information about the phases or adaptive strategies are needed to choose the right phase angles. In Section \ref{sec:phase_sensing}, we shall show that the PA protocols are adequate for achieving quantum advantages over any classical strategies.

\subsection{Correlation-to-displacement conversion} 
\label{sec:CtoD}

\begin{figure}[t]
\includegraphics[width=0.5\textwidth,trim=2 2 2 20,clip]{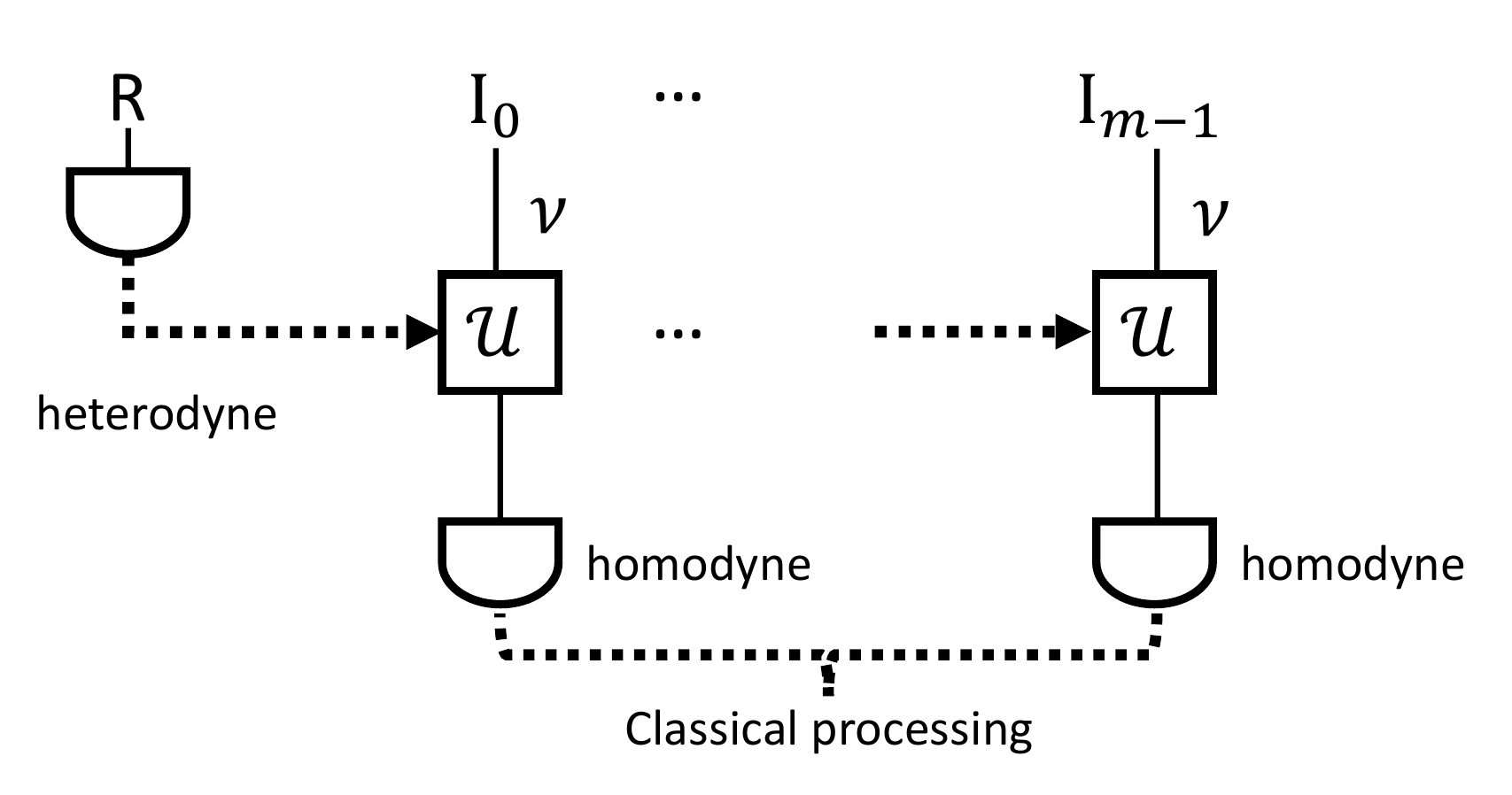}
\caption[]{Schematic of the CtoD measurement protocol. The procedure is comprised of three operations: (i) heterodyne detection on $\nu$ returning mode (ii) $\nu$-mode passive operations $\map U$ based on the heterodyne measurement results (iii) local homodyne detection and post-processing.  
}\label{fig:CtoD}
\end{figure}

In addition to the PA receiver network, one can consider the CtoD conversion protocol (see details in Appendix \ref{APP: QI-setup}), which involves the following steps: 
\begin{enumerate}
\item [(Q1)] Upon receiving the return, perform  \emph{heterodyne measurement} described by the positive operator-valued measure (POVM) $\{ |\chi \>\<\chi |/\pi  \}$ on the returning mode, where $|\chi\>$ refers to the coherent state with complex amplitude $\chi$. The measurement outcome can be described by the vector $\bs x=2[\text{Re}(\chi),\text{Im}(\chi)]^{\rm T}$, which satisfies the distribution:
\begin{align}\label{C2D-prob}
p(\bs x )&= [4 (N_{\rm B}'+1) \pi]^{-1}\exp\left[-[4(N_{\rm B}'+1)]^{-1}|\bs x |^2\right].
\end{align}
Conditioned on the measurement result $\bs x$,
the remaining idlers will have the mean and covariance matrix
\begin{align}\label{C2D-Dis-Cov}
\begin{cases}
\bs \xi_\chi&= [2(N_{\rm B}'+1)]^{-1} \bs S(\mathbb I_m\otimes \bs x)
,\\
\bs V_\chi&=(2N_{\rm S}+1)\mathbb I_{2m}-[2(N_{\rm B}'+1)]^{-1}\bs S\bs S^{\rm T},
\end{cases}
\end{align}
where  $\bs S=\left(S_0^{\rm T},\cdots, S_{m-1}^{\rm T}\right)^{\rm T}$ is a $2m\times 2$ matrix. Considering the multi-mode nature of the return, we denote the heterodyne measurement result as $\chi_n$ for each $n=0,\cdots,\nu-1$-th mode, where $\nu$ is the total number of modes.
\item [(Q2)] Implement identical $\nu$-mode passive operations $V_{\{|\bs x_n|\}}$ on each of the idler (containing $\nu$ modes) to align the displacements $\{\bs \xi_{\chi_n}\}$ and to concentrate all displacements of $m\nu$ modes into a single $m$-mode state (see details in Appendix \ref{APP:Qrepetition}. ) 
\item [(Q3)] Perform \emph{homodyne measurement} $\{|q\>\<q|\}$ on each of the $m$ concentrated idlers, where $|q\>$ refers to the eigenstate of the position operator $\hat q:=a+a^\dag$.  
\end{enumerate}
The diagram of the CtoD protocol is illustrated in Figure \ref{fig:CtoD}. The protocol is based on the CtoD conversion~\cite{shi2022fulfilling,shi2024optimal}, and is a direct generalization of the heterodyne-homodyne version of CtoD receiver~\cite{reichert2023quantum}. In the following parts of this paper, we will evaluate the quantum advantages of the PA and CtoD protocols in the context of multiple-phase sensing and pattern classification. 
\\[-8pt]

\section{Multiple-phase sensing}
\label{sec:phase_sensing}

With the receivers in hand, now we examine the performance of the QI network in multiple-phase sensing. We begin with the PA receiver in Section \ref{SEC:OPA_protocol}. Denote the PA gain of the pPCR as $g$; For sPCR, we also choose uniform PA gains and denote it as $g$, while the actual value of $g$ is optimized separately in sPCR and pPCR. Thanks to the Gaussian nature of the quantum states and operations involved (see Appendix \ref{APP:OPA}), we obtain the following theorem.
{\theo \label{theo2}\textbf{(PA network for multiple phase sensing)} In the QI network, the PA protocol can achieve the Fisher information matrix ($\nu\gg1$):
\begin{align}
&\map F^{\rm pa}_{\theta_j,\theta_k}=\nonumber \\
&2 \nu \cdot  \frac{\partial \bs b_j}{\partial \theta_j} \frac{\partial \bs b_k}{\partial \theta_k}\left[\bs a_j^{-1} \delta_{jk} -\frac{\left(\bs a_j\bs a_k\right)^{-1}\bs b_j\bs b_k}{1+\bs b^T\mathbf{diag}(\bs a)^{-1}\bs b}\right], \label{eq15}
\end{align}
where the coefficients are $\bs a_j=f_j ({g-1})  (N_{\rm B}'+1)(2N_{\rm S}+1) +N_{\rm S} $ and $\bs b_j=\sqrt{2f_j(g-1)   N_{\rm S}(N_{\rm S}+1)\eta_j/m_{\rm re}}\cos\theta_j$, $f_j=1/m$ for pPCR and $f_j=g^j$ for sPCR. %for pPCR. For sPCR, the coefficients are $\bs a'_j=g^j(g-1) (N_{\rm B}'+1)(2N_{\rm S}+1) +N_{\rm S} $ and $\bs b_j'= \sqrt{2g^j(g-1)  N_{\rm S}(N_{\rm S}+1)\eta_j/m_{\rm re}}\cos\theta_j$. %with  $\{\omega_j':=\left({\rm Re}\sqrt{\omega_{j}}\right)^2\}$ referring to the adjusted weights. 

The corresponding rWMSE $\epsilon_{\bs \theta}=\sqrt{\Tr \left[{(\map F^{\rm pa})}^{-1}\right]/m}$ to the leading order can be obtained as
\begin{align}
\epsilon_{\bs\theta}&=\sqrt{\sum_{j=0}^{m-1} \frac{m_{\rm re}[f_j(g-1)(N_{\rm B}'+1)(2N_{\rm S}+1)+N_{\rm S}]}{2m\nu\left[2 f_j(g-1)N_{\rm S}(N_{\rm S}+1)\eta_j\cos^2\theta_j\right]}}\nonumber \\
&+\mathcal O \left(\sqrt{\frac{ N_{\rm S}}{m_{\rm re}\nu  N_{\rm B}}}\right).
\label{eq9}
\end{align} 
}\\[-6pt]

A detailed proof of Theorem \ref{theo2} can be found in Appendix \ref{APP:OPA}. In the above result, the PA gain is not specified. Indeed, one can optimize the sensing performance by tuning the values of gain. We observe that the ideal amplification rate $g$ varies between pPCR and sPCR. In pPCR, the best value of $g$ is close to two, while in sPCR, it is close to one.

%\QZ{***is there a reason you want to show Fisher information matrix for PCR but not for CtoD?***}\XZ{For PCR, we have a simple and exact expression of Fisher information matrix. For CtoD, we only have bounds and complicated Fisher information matrix.  }

For the CtoD measurement approach, exact expression of the Fisher information matrix is complicated. On the other hand, we can obtain the rWMSE expression asymptotically.

{\theo\label{theo3}\textbf{(CtoD  conversion for multiple phase sensing)} In the QI  network for multiple phase sensing, the CtoD protocol achieves the rWMSE
\begin{align}\label{wrmse}
\epsilon_{\bs \theta}=&
\sqrt{\frac{m_{\rm re}(N_{\rm B}'+1)(2N_{\rm S}+1)}{4m\nu N_{\rm S}(N_{\rm S}+1)}\sum_{j=0}^{m-1} \eta_j^{-1}} \nonumber \\
&+ \mathcal O\left(\sqrt{\frac{ N_{\rm S}}{m_{\rm re}\nu N_{\rm B}}}\right).
\end{align}
%The second and third steps of CtoD are optimal conditioned on the first step and the assumption $N_{\rm S}< N_{\rm B}$.
%In the estimation of the average phase $\overline \theta=\sum_j \sqrt{\omega_j'\eta_j}\theta_j$, the  root mean-square-error (RMSE) achievable by the CtoD method is: $\epsilon_{\overline \theta}=\epsilon_{\bs\theta}/\sqrt{\sum_j\left(\omega_j'\eta_j\right)^{-1}}$.
%$\epsilon_{\overline \theta}=\sqrt{(N_{\rm B}'+1)(2N_{\rm S}+1)/[4m\nu N_{\rm S}(N_{\rm S}+1)]}+\mathcal O\left(\sqrt{{ N_{\rm S}}/(m\nu N_{\rm B})}\right)$. 
}

\begin{figure}[t]
\includegraphics[width=0.47\textwidth,trim=1 1 1 1,clip]{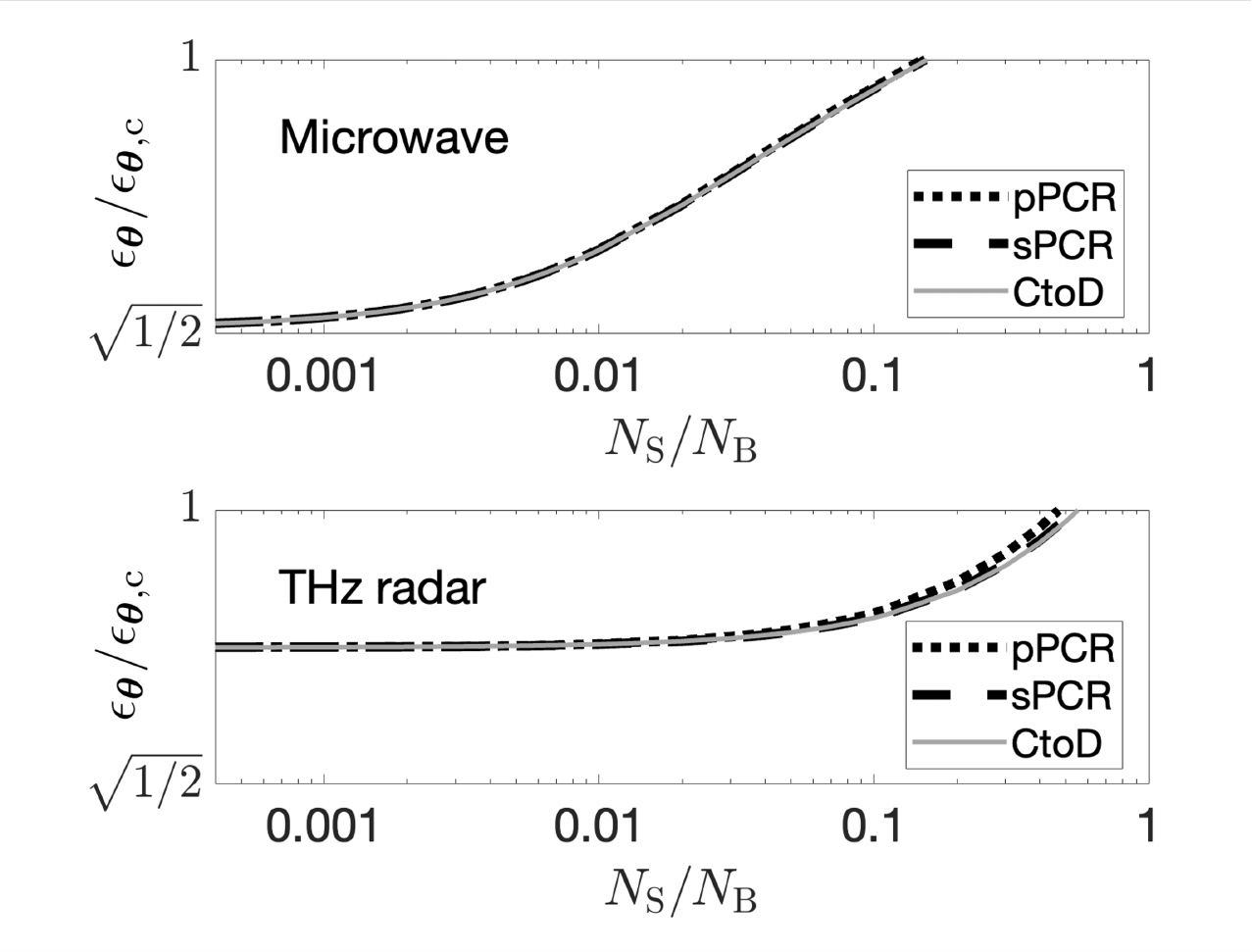}
\caption[]{rWMSE ratio of phase estimation regarding the SNR. Here we precisely evaluate two cases: (i) microwave Radar with $N_{\rm B}=32$;(ii) THz Radar with $N_{\rm B}=0.6$, with  $m=50$ transmitters with $\nu=5000$ rounds of experiment. For PA receivers, the amplification rate is $g\sim 2m$ for pPCR and $g\sim 2$ for sPCR. Its error refers to the minimum error of all phase values. For CtoD protocol, the error is calculated for $\theta_j\to 0$ and  $\eta_j\sim0.5$. 
}\label{fig:NS_NB}
\end{figure}

A concrete proof of Theorem \ref{theo3} is shown in Appendix \ref{APP:advantage_phase_sensing}. To thoroughly understand the trend of the precise rWMSE obtained by the QI network in terms of the signal and noise brightness, we consider the two application scenarios introduced in Section~\ref{sec:application}, where the noise $N_{\rm B}=32$ for a W-band radar and $N_{\rm B}=0.6$ for a THZ radar. Then we tune the signal brightness $N_{\rm S}$ and evaluate the phase sensing rWMSE error $\epsilon_\theta$ for pPCR, sPCR and CtoD in QI network. In Fig.~\ref{fig:NS_NB}, we plot the ratio of the quantum rWMSE (Eqs.~\eqref{eq9} and~\eqref{wrmse}) over the classical rWMSE (Eq.~\eqref{cla_rWMSE}) versus the ratio $N_{\rm S}/N_{\rm B}$. It is shown that, as the input signal brightness grows, the quantum advantage tends to vanish as expected. On the other hand, given the low-brightness limit $(N_{\rm S}\ll 1)$, the ratio between the rWMSEs will converge to a constant factor~\cite{sanz2017quantum,shi2024optimal}---a factor of two advantage in terms of the variance. %At the same time, we see that sPCR and CtoD receivers have better performance in the non-asymptotic region of $N_{\rm S}$, when compared with the pPCR.

Remarkably, Theorems \ref{theo2} and \ref{theo3} extend upon earlier approaches of quantum illumination~\cite{lloyd2008enhanced,tan2008quantum,sanz2017quantum,shapiro2020quantum,karsa2023quantum} by introducing an efficient measurement design for the QI network in the typical parameter region of $N_{\rm S}<N_{\rm B}$. In particular, the errors shown in Eqs. (\ref{eq15}) and (\ref{wrmse}) converge to the same value in the weak signal limit of $N_{\rm S}\ll N_{\rm B}$ (see Figure \ref{fig:NS_NB}). Further, Eq. (\ref{wrmse}) achieves the quantum limit for single parameter estimation  \cite{shi2020practical,shi2022fulfilling}, thus being tight in the scenario when only one phase is unknown. In addition, it is shown in Appendix \ref{APP:advantage_phase_sensing} that the second and third steps of the CtoD method are optimal, conditionally on the choice of heterodyne measurement in its first step. 

Note that Eqs. (\ref{eq9}) and (\ref{wrmse}) approach a scaling of $\mathcal O(\sqrt{m_{\rm re}/\nu})$. %even if we assume the interference amplitudes have a uniform distribution, i.e. $\omega_j\sim \mathcal O(1/m)$. 
This is caused by the fact that information about each phase is vanishing as its corresponding quantum amplitude decreases. To resolve this issue, we can alternate the parameter of interest to the average phase $\overline \theta$, which induces an error of scaling $\mathcal O(\sqrt{1/(m\nu)})$. Thus, it is possible to achieve a finite error with an arbitrarily high number of transmitters, even when the number of transmitters $m$ far exceeds the number of experiments $\nu$.

\iffalse 
\begin{figure}[!ht]
\includegraphics[width=0.5\textwidth,trim=1 1 1 1,clip]{rWMSE.pdf}
\caption[]{Estimation error of illumination networks.  (Upper plot) microwave Radar with $N_{\rm B}=32$;  (Lower plot) THz Radar with $N_{\rm B}=0.6$. The number of transmitters is $m=20$. The expected values of parameters and the interference weights are $\eta_j=0.5$,  $\theta_j\sim 0$, and $\omega_j=1/m$ for $j=0,\cdots,m-1$.
}\label{fig:rWMSE}
\end{figure}
\fi

In addition to estimating the phase of each mode, the QI network can also be used to estimate the reflectivity (assuming knowledge of the phases).
Given that the PA and CtoD have the same leading order of rWMSE, by changing parameters of interest, we have the following Corollary: 
%Given that the CtoD protocol can achieve better precision over the OPA methods, we investigate the sensing of multiple reflectivity ratios with CtoD and have the following Corollary: 
{\cor \label{corr1}\textbf{(Reflectivity sensing)} In QI reflectivity sensing for $\{\eta_j\}$, the achievable rWMSE is:
\begin{align}
\epsilon_{\bs \eta }\le  2\,\epsilon_{\bs\theta}\sqrt{\frac{\sum_{j} \eta_j}{\sum_{k} \eta_k^{-1}}}\,+ \mathcal O\left(\sqrt{\frac{  N_{\rm S}}{m_{\rm re}\nu N_{\rm B}}}\right). 
\end{align} 
}\\[-0.2em]
Corollary \ref{corr1} can be quickly verified by substituting parameters of interest by reflectivity ratios when computing QFIMs (see Appendix \ref{APP:advantage_phase_sensing}). \\[-0.2em]

\noindent \textbf{Remark 1:  ($(m+1)$-mode vs single-mode output states) } The QI network produces a Gaussian state with $(m+1)$ modes, using $m$ copies of TMSV input states (see Equation (\ref{CoV})). The zero-mean state's properties are defined by its $2m$-by-$2m$ covariance matrix, which in turn allow for the estimation of $m$ independent phases. In contrast, classical correlations can only be produced by classical mixtures of states, which does not help in achieving the bound in Eq. (\ref{cla_rWMSE}). In addition, if the input state is pure, the corresponding output state will be a single-mode displaced thermal state, in which the information of $m$ independent phases can only be determined by its displacement, with only two independent degree of freedom. The limited degree of freedom cannot allow the independent extraction of $m$ parameters. In Appendix \ref{APP:benchmark_phase_sensing}, we show that the rWMSE for arbitrary classical estimation protocol is subjected to $\epsilon_{\bs\theta,{\rm c}}= \sqrt{{2\nu} E_{{\rm 1-c}}/m +\left(1-{2\nu}/{m}\right) {\pi^2}/{3}}$ in the case $\nu<  m /2 $, where $E_{\rm 1-c}$ refers to the minimum mean-square-error achievable via a single-shot measurement of the output state from classical illumination networks.\\[-0.5em]

\noindent \textbf{Remark 2:  (PA receiver vs CtoD method)} Here we address different challenges of experimental implementation of both protocols in the microwave frequency region. While the PA receiver only requires Gaussian operations and photo detection, the implementation of PA receiver requires direct interaction between the noisy return with the idler stored in the fridge (see Fig.~\ref{fig:OPA_PCR}), which may induce additional loss and noise to the idler system. As shown in Fig.~\ref{fig:CtoD}, the CtoD receiver avoids the direct interaction between the noisy return and the idler stored in the fridge and only relies on classical feed-forward to connect the separable measurements on the signals and idlers. Furthermore, the CtoD receiver consistently outperforms all other PA receivers in terms of estimate error.

%\noindent \textbf{Remark 2 (Asymptotic vs non-asymptotic quantum metrology)} With finite trials of experiments, one has to encode 

\section{Pattern classification}
\label{sec:pattern}

Besides multi-parameter estimation, hypothesis testing between different targets is one of the first applications of QI, particularly in determining the presence or absence of a target \cite{lloyd2008enhanced,tan2008quantum}. In the setting with a QI network, the hypotheses can be described by the change of possible values of reflectivity: $\text{(I)\ \ \ } \bs \eta=\bs \eta^{(0)}$, and $
\text{(II)\ \ } \bs \eta=\bs \eta^{(1)}$, 
where $\bs \eta^{(h)}=(\eta_0^{(h)},\cdots,\eta_{m-1}^{(h)})^{\rm T}$ for $h=0,1$. Given multiple copies of the $(m+1)$-mode signal-idler state, the CtoD scheme will generate conditional states at $j$-th mode, whose displacement depends on the reflectivity $\eta_j$ and the heterodyne measurement results $\{{\bs x_n}\}$. By conducting passive operations on the $\nu$ copies of each mode, it is possible to produce $\nu$ identical $m$-mode states only based on the knowledge of the heterodyne measurement results %\QZ{why is it identical copies? the measurement results of the CtoD is random, what is $\bm x_n$?} 
({see detailed proof in Appendices \ref{APP:Qrepetition} }and \ref{APP:PC}): $\rho_{\bs \eta^{(h)},{\rm hp},\{\bs x_n\}}=  \rho_{\bs \eta^{(h)},\overline{\bs x}}^{\otimes \nu}. $ Therefore, if we quantify the performance of pattern classification by the error probability:
$p_{\rm hp} \left(\rho_{\bs \eta^{(0)},{\rm hp}},\rho_{\bs \eta^{(1)},{\rm hp}}\right) =1-\max_{\{\hat \Pi_{h} \}} \Tr\left[\hat \Pi_{h,\overline{\bs x}}\rho_{\bs \eta^{(h)},{\rm hp}}\right]/2$ with ${\{\hat \Pi_{h} \}}$ being an arbitrary $(m\nu)$-mode measurement, the following two theorems are given: 

% \begin{figure}[t]
% \includegraphics[width=0.4\textwidth,trim=2 1 2 1,clip]{rWMSE_m.pdf}
% \caption[]{Average-phase estimation error in terms of transmitter numbers. Here we numerically simulate the Microwave Radar with photon numbers $N_{\rm B}=32$, $N_{\rm S}=0.5$, and parameters $\eta_j\sim 0.5$ and $\theta_j\sim0$. The maximal modes is $m_{\rm re}=125$. Number of experiments is $\nu=30$. The orange dashed line is obtained by the first term of Eq. (\ref{eq13}) in a scaling $\mathcal O(1/\sqrt m)$. For the PA protocols, the amplification ratio $g\sim 2m$ for pPCR and $g\sim1.2$ for sPCR. Here the lines of pPCR, sPCR, and CtoD basically coincides with the gray dotted line.  
% }\label{fig:rWMSE_m}
% \end{figure}

\begin{figure}[t]
\includegraphics[angle=0,width=0.6\textwidth,trim=130 90 0 90,clip]{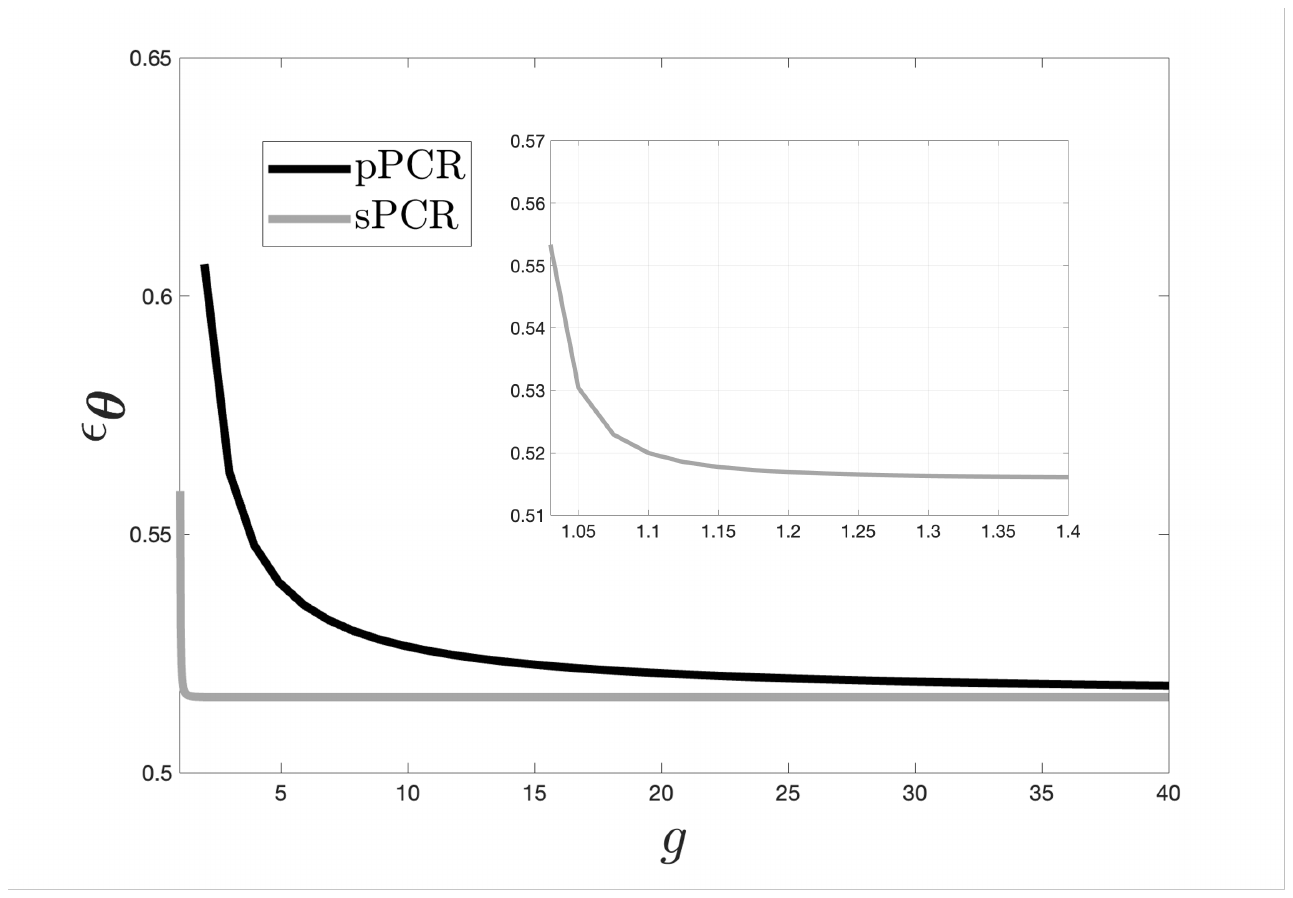}
\caption[]{Phase-estimation errors of PA receiver networks regarding amplification ratios. Here we numerically simulate the microwave Radar with photon numbers $N_{\rm B}=32$, $N_{\rm S}=0.5$, and parameters $\eta_j\sim 0.5$ and $\theta_j\sim0$. The maximal modes is $m_{\rm re}=120$. Number of transmitters is $m=50$. Experiments rounds is $\nu=2\cdot 10^4$. 
}\label{fig:OPA_g}
\end{figure}

{\theo\label{pro2}\textbf{(Benchmark for pattern classification)} When the reflectivities are unknown or time-varying, which prevents the concentration of power in some probe modes, the minimal error probability for multiple-pattern classification is: 
\begin{align}
p_{\rm ci} \sim&\frac 1 2 \exp\left\{-\frac {\nu N_{\rm S}\left|\sum_{j=0}^{m-1}\left(\sqrt{\eta_{ j}^{(0) }}-\sqrt{\eta_{j}^{(1) }}\right)\right|^2} {m_{\rm re}\left(\sqrt{N_{\rm B}+1}+\sqrt{N_{\rm B}}\right)^2} \right\}.
\label{pci}
\end{align}}
A concrete proof of Theorem \ref{pro2} can be found in Appendix \ref{APP:PC}. It is shown that the benchmark can be achieved by preparing a statistic mixture of coherent states as input  \cite{weedbrook2012gaussian,shapiro2020quantum}.

On the other hand, we have the following theorem if entangled probes are allowed: 
{\theo\label{pro1}\textbf{(Quantum limit for pattern classification)} The QI network that satisfies the conditions that $N_{\rm S}=\mathcal O(1)\ll N_{\rm B}$ can achieve the error probability: 
\begin{align}\label{eq8a}
p_{\rm qi}
\sim &\nonumber  \frac 1 2 \exp\left[- \frac{\nu N_{\rm S}(N_{\rm S}+1)\sum_{j=0}^{m-1} \left(\sqrt{\eta_j^{(0) }}-\sqrt{\eta_j^{(1) }}\right)^2}{m_{\rm re}(N_{\rm B}'+1)\left(\sqrt{N_{\rm S}+1}+\sqrt{N_{\rm S}}\right)^2}\right]\\&+ \mathcal O\left(\frac{m^2\nu}{m_{\rm re}^2N_{\rm B}^2}\right).
\end{align} }
A concrete proof of Theorem  \ref{pro1} can be found in Appendix \ref{APP:PC}.

\noindent \textbf{Remark 3:  (Quantum advantage vs disadvantage)} We note that the CI and QI error probabilities in Eq.~\eqref{pci} and Eq.~\eqref{eq8a} have different dependence on the reflectivities---the classical case has amplitude summed and then square while the quantum case has amplitude squared and then summed. Such a difference comes from the interference in Eq.~\eqref{eq2}: in the classical case, Eq.~\eqref{eq2} will directly combine coherent state in amplitudes; while in the quantum case the multiple idlers are obtained in weakly thermal coherent states and they will only be combined in energy even if one further applies beamsplitter to concentrate all coherent states. Similar scaling difference can also be identified for single parameter estimation (see Appendix~\ref{app:average_phase_dis}). Remarkably, this phenomenon does not appear in multi-parameter estimation. In simple terms, the reason is that Eq.~\eqref{eq9s} excludes the off-diagonal elements of the inverse of Fisher information matrices, resulting in the rWMSE being unaffected by how parameters are collectively encoded in states.  
%Thereby, the form of combining parameters is discarded due to the calculation with derivatives (see  Eq. (\ref{eqd1}) in Appendix \ref{APP:QFI}). More explicitly, the intermediate CtoD procedure approximately has a   conditional displaced thermal state (Eq. (\ref{C2D-Dis-Cov})) with a probability (Eq. (\ref{C2D-prob})), while classical illumination networks generate another displaced thermal state (Eq. (\ref{eqa1}) in Appendix \ref{APP:benchmark_phase_sensing}). As shown in Appendix \ref{APP:QFI}, the QFIM entry regarding parameters $\phi_j$ and $\phi_k$ for displaced thermal state with an average thermal photon number $N_{\rm T}$ and displacement $\bs d$ is: 
%\begin{align}F_{\phi_j,\phi_k}&\sim \frac{1}{2N_{\rm T}+1}\left(\frac{\partial \bs d}{\partial \phi_j}\right)^T\left(\frac{\partial \bs d}{\partial \phi_k}\right).\end{align}
%Thus, the combination among parameters in displacement (eg. Eq. (\ref{eqa1}) in Appendix \ref{APP:benchmark_phase_sensing}) is omited by taking derivatives. 

Finally, Eq. (\ref{eq8a}) achieves the optimal error probability for hypothesis testing with a single parameter \cite{shi2022fulfilling}. The error exponent of the QI network has an overhead as follows: 
\begin{align}
\frac{\ln(2\,p_{\rm qi})}{\ln(2\,p_{\rm ci})}&\nonumber =\frac{\left(\sqrt{N_{\rm B}+1}+\sqrt{N_{\rm B}}\right)^2(N_{\rm S}+1)}{\left(\sqrt{N_{\rm S}+1}+\sqrt{N_{\rm S}}\right)^2(N_{\rm B}'+1)}\\
&\times \frac{\sum_{j=0}^{m-1} \left(\sqrt{\eta_j^{(0) }}-\sqrt{\eta_j^{(1) }}\right)^2}{\left|\sum_{j=0}^{m-1}\left(\sqrt{\eta_{ j}^{(0) }}-\sqrt{\eta_{j}^{(1) }}\right)\right|^2}
\\
&\simeq 4 \frac{\sum_{j=0}^{m-1} \left(\sqrt{\eta_j^{(0) }}-\sqrt{\eta_j^{(1) }}\right)^2}{\left|\sum_{j=0}^{m-1}\left(\sqrt{\eta_{ j}^{(0) }}-\sqrt{\eta_{j}^{(1) }}\right)\right|^2},\label{eq18}
\end{align}
where in the second approximation we have considered the $N_{\rm B}\gg1$ and $N_{\rm S}\ll1$ limit.
    When the number of transmitters $m=1$, the above results recovers the general hypothesis testing result in Ref.~\cite{shi2024optimal} and provides a factor of ${\ln(2\,p_{\rm qi})}/{\ln(2\,p_{\rm ci})}\simeq 4$ advantage in error exponent (6 dB). In typical cases where the differences  $\left\{\sqrt{\eta_{ j}^{(0) }}-\sqrt{\eta_{j}^{(1) }}\right\}$ has values $\pm c$ with equal probabilities (representing absent or present), there will be an exact 6-dB advantage in error exponent as the denominator of Eq. (\ref{eq18}) will be proportional to $m$ \cite{lawler2010random}.

% \begin{figure}[!ht]
% \includegraphics[width=0.5\textwidth,trim=2 0 2 0,clip]{pattern classification}
% \caption[]{Numarical simulation of the error probability obtained by the QI network and classical illumination methods. Here the black dashed lines are the error probability achieved by the QI network, while the grey dashed lines are the error probability achieved by classical probe states. We consider a microwave Radar with $N_{\rm S}=1$ and $N_{\rm B}=32$. There are $m=50$ transmitters in total, used to estimate $m$ random reflectivities. Here the interference weights and the phases are assumed to have the value $\omega_j=1/m$ and $\theta_j\sim0$, respectively.  
% }\label{fig:Pattern_Classification}
% \end{figure}

%In the situation where the experimenter knows which mode will induce the largest difference between the reflectivities, it can concentrate all the $mN_{\rm S}$ photons to one mode to attain a better performance in the hypothesis testing. Nevertheless, the QI network can also use all the $mN_{\rm S}$ photons in a single TMSV mode, and achieve a 6dB advantage as in the single-parameter hypothesis testing provided the condition $mN_{\rm S}\ll 1$. 

\section{Discussion: Potential advantage in the optical region}\label{sec:potential_opt}
Optical region typically there is no-go theorem on quantum advantage when it is very lossy \cite{tan2008quantum,lloyd2008enhanced,shapiro2020quantum,karsa2023quantum}. Here we present results in the non-asymptotic region that may bring hope to quantum advantage in lossy optical sensing.
In particular, this non-asymptotic estimation situation arises when the transmitter number is comparatively larger than the number of experiment rounds. On this account, the following Theorem can be given: \\[-0.5em]

{\theo \label{theo1.1}\textbf{(Non-asymptotic benchmark for phase sensing $(\nu< \lceil m /2  \rceil )$)} The classical benchmark for multiple phase sensing with the uniform distribution has the following non-asymptotic bounds: 
\begin{align}\label{nonasymLow}
\epsilon_{\bs \theta,{\rm c}}&\ge \min_J \sqrt{\frac{J}{m}\frac{\min_j E_{\rm Baye}(\rho_{\bs\beta_h(j)},\theta_j)}{\nu'} +\left(1-\frac{J}{m} \right) \frac{\pi^2}{3}}, \\ 
\epsilon_{\bs \theta,{\rm c}}&\le \min_{J'} \sqrt{\frac 1 m \sum_{k=0}^{J' -1} \frac{E_{\rm c,homo,k}}{\nu'}+\left(1-\frac {J'} m  \right)\frac{\pi^2}{3}}.\label{nonasymUp} 
\end{align}
where $E_{\rm Baye}(\rho(\bs\theta),\theta_j) $ is a Beyesian variance \cite{helstrom1969quantum,personick1971application,yuen1973multiple,rubio2020bayesian}, $\{E_{\rm c,homo,k}\}$ is the mean-square-error achievable by preparing coherent state probes and performing homodyne measurement for $J=\lfloor 2 \nu/\nu'\rfloor(/ J'=\lfloor\nu/\nu'\rfloor)$ phases with $\nu'=1,2,\cdots,2\nu/J(\nu/J')$.  
}\\[-0.5em]

\noindent A concrete demonstration of Theorem \ref{theo1.1} is shown in Appendix \ref{app:non-asymptotic}. In simple terms, when the value of $\nu$ is much less than $m$, the rWMSE converges to a constant. This phenomenon is caused by the limitation that coherent state probes can only have two independent phases. Additionally, a statistical mixture of coherent states will induce a biased measurement results at each possibility, thus leaving other parameters at random guess. On the other hand, the QI network with CtoD measurement protocol will generate a $(m+1)-$mode output state, providing us with the opportunity to estimate each of $m$ independent parameters for enough experiment rounds without random guess (see Section \ref{sec:phase_sensing}). 

\begin{figure}[t!]
\includegraphics[width=0.35\textwidth,trim=2 10 2 2,clip,angle =270]{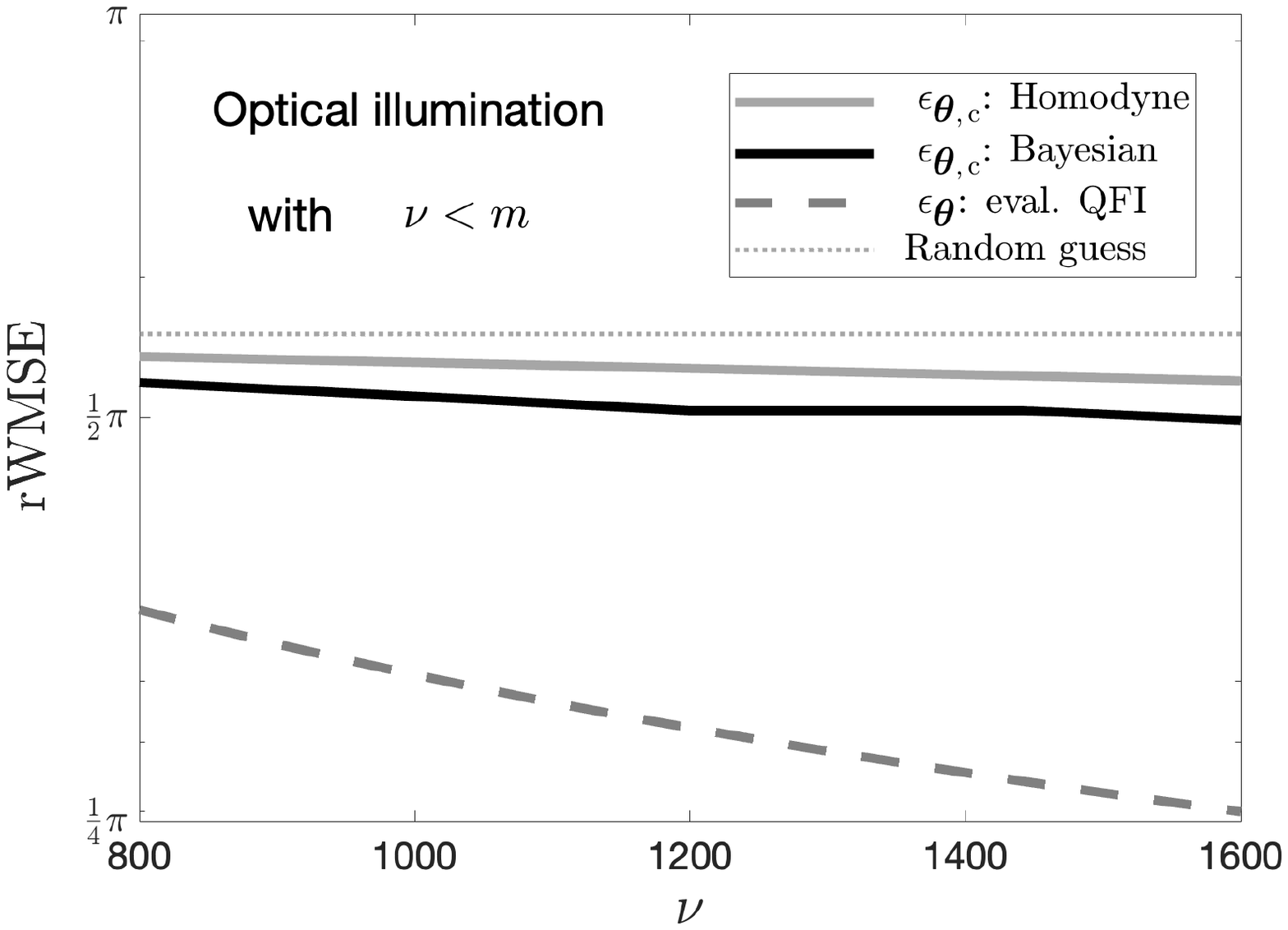}
\caption[]{Performance in non-asymptotic QI phase sensing. Here we numerically evaluate the optical scenario with $N_{\rm S}=200$, $N_{\rm B}=0.01$, $m=1600$ effective transmitters, $m_{\rm re}=2000$, and transmissivity ratios $\eta_j\sim 0.7$. The classical benchmarks $\epsilon_{\bs \theta,{\rm c}}$ is evaluated by assuming that the whole energy $mN_{\rm S}$ is concentrated in a single probe in each experiment trail. Given a fixed experiments round number $\nu$, we optimize the phase number $J,J'= \nu/\nu'$ to be estimated while leaving the other phases at random guess. The upper bound of benchmark is evaluated by $\nu'$-shot homodyne measurement with MSE $E_{\rm c,homo,k}\approx 6.31/\nu'$. The lower bound of benchmark is evaluated by the Bayesian bound with $E_{\rm Baye}=3.53/\nu'$. The non-asymptotic Bayesian bound for QI network is evaluated by the asymptotic Cram\'er-Rao bound obtained with parametric amplifier networks (see Theorem \ref{theo2} of Section \ref{sec:phase_sensing}). The explicit calculation of the cases where two classical probes are used, i.e. $J=2\nu/\nu'$ is an open problem as it induces negative Beyasian bound in this case. In addition, the derivation of the non-asymptotic performance of  QI network remains unresolved.  }
\label{fig:non_asym}
\end{figure}

Interestingly, this disadvantage of non-asymptotic classical estimation protocols might be present in the optical region, unlike the conventional illumination situation detecting a single parameter  \cite{shi2022fulfilling}. Here, we illustrate the performance of optical illumination networks in the scenario $\nu<m$ in Fig. \ref{fig:non_asym}. Specifically, we evaluate the lower bound and upper bound of the classical benchmark by considering Eq. (\ref{nonasymLow}) and Eq. (\ref{nonasymUp}), respectively. Due to the difficulty of computing non-asymptotic bound of rWMSE with correlated measurement data (see Section \ref{sec:phase_sensing}), we evaluate the non-asymptotic error with asymptotic Cram\'er-Rao bound in Fig. \ref{fig:non_asym} and leave the explicit proof of this potential open for future approaches.

\section{Conclusions} 
\label{sec:conclusions}

We investigated the metrological usefulness and measurement design in the quantum illumination network, where a transmitter array and a single receiver antenna are used. We proved an advantage in multiple-parameter sensing and hypothesis testing. Specifically, we analytically computed the minimum rWMSE achievable by both the QI network and arbitrary classical strategies for the typical scenarios where the probe photon number is smaller than that of the background noise. We show that for a significant range of probe photon numbers, the rWMSE achieved by either a PA or CtoD measurement protocol has a favorable value. In contrast, all classical strategies are subjected to a larger error when the signal is comparatively smaller than the noise. Further, we extend the discussion to pattern classification of the reflectivity pattern. We show that when each transmitter is constrained to a fixed number of photons, the QI network can achieve the six-decibel advantage as in the single-parameter hypothesis testing case. As explored in Ref.~\cite{zhuang2022ultimate}, we expect the six-decibel advantage can lead to large resolution advantage in the threshold region of parameter estimation.

Finally, we point out a few future directions. In the QI network, we have only considered a single receiver antenna. An array of antenna such as that in the multiple-input and multiple-output (MIMO) channel scenarios can enable further enhancement of signal-to-noise ratios. While we considered multi-parameter estimation, it is an open question how such advantages can be generalized to more general measurement settings, such as tomography and learning.

\noindent\emph{Acknowledgments.---} This project was supported by the National Science Foundation CAREER Award CCF-2142882,  the National Science Foundation (NSF) Engineering Research Center for Quantum Networks Grant No. 1941583, Office of Naval Research (ONR) Grant No. N00014-23-1-2296, National Science Foundation OMA-2326746, and the Defense Advanced Research Projects Agency (DARPA) under Young Faculty Award (YFA) Grant No. N660012014029. QZ also acknowledge supports from Halliburton Technology.

%The  parameter estimation of this particular case have been investigated for identical phase \cite{shi2020practical} and reflectivity \cite{sanz2017quantum}. 

\bibliographystyle{apsrev4-1}
\bibliography{reference}

\begin{widetext}

\appendix

\section{Benchmark for phase estimation}\label{APP:benchmark_phase_sensing}

In this Appendix, we shall demonstrate the estimation error limit that can not be surpassed by any classical schemes. 

\subsection{Set-up}

Without losing the generality, let us consider a situation in which the experimenter is constrained to signal-idler in mixtures of coherent states  \cite{serafini2017quantum,shapiro2020quantum}: 
\begin{align}\label{eqa1}
\rho_{{\rm c}}=&\int \d \bs \beta \,p(\bs \beta) \bigotimes_{j=0}^{2m-1} |\beta_j\>\<\beta_j|\otimes |0\>\<0|^{\otimes (m_{\rm re}-m)},
\end{align}
where $\bs \beta=(\beta_0,\cdots,\beta_{2m-1})$ refers to the vector of complex amplitudes and $\{p(\bs \beta)\}$ denotes an arbitrary probability distribution. 
The energy constraint can be expressed as:
\begin{align}\label{eqa2}
\sum_{j=0}^{m-1}\int \d \bs \beta \,p(\bs \beta) |\beta_j|^2\le m\, N_{\rm S}.
\end{align}

Then, the output state of a multi-parameter illumination process becomes: 
\begin{align}
\rho_{\rm ci}&=\map Q_{\bs \eta,\bs \theta} \left(\rho_{{\rm c}}\right)\\
&= \int \d \bs \beta\, p(\bs \beta) \map Q_{\bs \eta,\bs\theta} \left(\bigotimes_{j=0}^{2m-1} |\beta_j\>\<\beta_j|\otimes |0\>\<0|^{\otimes (m_{\rm re}-m)}\right), \\
&= \int \d \bs \beta\, p(\bs \beta) D\left(\sum_{j=0}^{m-1}\sqrt{\omega_{j}\eta_{j}}e^{-i\theta_{j}}\beta_{j}\right)\rho_{N_{\rm B}}D\left(\sum_{j=0}^{m-1}\sqrt{\omega_{j}\eta_{j}}e^{-i\theta_{j}}\beta_{j}\right)\bigotimes_{k=m}^{2m-1} |\beta_k\>\<\beta_k|,\label{eqa4}
\end{align}
where $\map Q_{\bs \eta,\bs \theta}$ is the quantum channel of illumination process, $\rho_{N_{\rm B}}$ refers to a thermal state with an average photon number $N_{\rm B}$, $\left\{\left.\omega_j>0,j=0,\cdots,m_{\rm re}\right|\sum_{j=0}^{m_{\rm re}-1}\omega_j=1\right\}$ are weights defined by linear operation $a_j\to \omega_j a_{\rm R}$ in consideration of $m$ signaling transmitters and $m_{\rm re}-m$ vacuum inputs, $\{\eta_j\}$ and $\{\theta_j\}$ are the parameters defined by Eq. (\ref{thermal-loss}) in the main text.  Without loss of generality, we adopt the simplification  $\omega_j=1/m_{\rm re}$ as in Eq. (\ref{eq2}) of the main text in estimating error scalings throughout all Appendices, as imhomogenuity can be absorbed into the different values of $\eta_j$'s.

\subsection{Lower bound of rWMSE}

Given the definition of parity for matrices $A\ge B\Longleftrightarrow A-B\ge 0$ and the convexity of quantum Fisher information matrix (QFIM)   \cite{takeoka2017fundamental,liu2020quantum,slaoui2022analytical}, we have the following bound: 
\begin{align}
F_{\bs \theta,{\rm c}}&=\max_{p} F_{\bs \theta}\left[\int \d \bs \beta\, p(\bs \beta) \map Q_{\bs \eta,\bs\theta} \left(\bigotimes_{j=0}^{2m-1} |\beta_j\>\<\beta_j|\otimes |0\>\<0|^{\otimes (m_{\rm re}-m)}\right)\right] \\
&\le \max_{p} \int \d \bs \beta\, p(\bs \beta) F_{\bs \theta}\left[\map Q_{\bs \eta,\bs\theta} \left(\bigotimes_{j=0}^{2m-1} |\beta_j\>\<\beta_j|\otimes |0\>\<0|^{\otimes (m_{\rm re}-m)}\right)\right] \label{c6}\\
&= \max_{p} \int \d \bs \beta\, p(\bs \beta)  \frac 1 {2N_{\rm B}+1}\left(C_{\bs \beta}\bs \eta_{\omega}\bs \eta_{\omega}^{\rm T} C_{\bs \beta}+S_{\bs \beta}\bs \eta_{\omega}\bs \eta_{\omega}^{\rm T} S_{\bs \beta}\right)\\
&=  \frac 1 {2N_{\rm B}+1} \max_{p} \frac{\int \d \bs \beta\, p(\bs \beta) C_{\bs \beta}\bs \eta_{\omega}\bs \eta_{\omega}^{\rm T} C_{\bs \beta}+S_{\bs \beta}\bs \eta_{\omega}\bs \eta_{\omega}^{\rm T} S_{\bs \beta}}{\int \d \bs \gamma\, p(\bs \gamma) \Tr \left(C_{\bs \gamma}\bs \eta_{\omega}\bs \eta_{\omega}^{\rm T} C_{\bs \gamma}+S_{\bs \gamma}\bs \eta_{\omega}\bs \eta_{\omega}^{\rm T} S_{\bs \gamma}\right)}\int \d \bs \xi\, p(\bs \xi) \Tr \left[C_{\bs \xi}\bs \eta_{\omega}\bs \eta_{\omega}^{\rm T} C_{\bs \xi}+S_{\bs \xi}\bs \eta_{\omega}\bs \eta_{\omega}^{\rm T} S_{\bs \xi}\right]\\
&\le  \frac {4mN_{\rm S}\max_j \omega_j\eta_j} {2N_{\rm B}+1} \max_{p} \frac{\int \d \bs \beta\, p(\bs \beta) C_{\bs \beta}\bs \eta_{\omega}\bs \eta_{\omega}^{\rm T} C_{\bs \beta}+S_{\bs \beta}\bs \eta_{\omega}\bs \eta_{\omega}^{\rm T} S_{\bs \beta}}{\int \d \bs \gamma\, p(\bs \gamma) \Tr \left[C_{\bs \gamma}\bs \eta_{\omega}\bs \eta_{\omega}^{\rm T} C_{\bs \gamma}+S_{\bs \gamma}\bs \eta_{\omega}\bs \eta_{\omega}^{\rm T} S_{\bs \gamma}\right]}, 
\end{align}
where $F_{\bs \theta}(\rho)$ refers to the QFIM for the state $\rho$ defined in Eq. (\ref{eq9s}),  $C_{\bs \beta}={\textbf{diag}}(2|\beta_{0}|\cos (\theta_0+{\text{arg}}(\beta_0)+{\text{arg}}(\omega_0)/2),\cdots,2|\beta_{m-1}|\cos (\theta_{m-1}+{\text{arg}}(\beta_{m-1})+{\text{arg}}(\omega_{m-1})/2))$, $S_{\bs \beta}={\textbf{diag}}(2|\beta_{0}|\sin (\theta_0+{\text{arg}}(\beta_0)+{\text{arg}}(\omega_0)/2),\cdots,2|\beta_{m-1}|\sin (\theta_{m-1}+{\text{arg}}(\beta_{m-1})+{\text{arg}}(\omega_{m-1})/2))$ and $\bs \eta_\omega = \left(\sqrt{\omega_{0}\eta_{0}},\cdots,\sqrt{\omega_{m-1}\eta_{m-1}} \right)^{\rm T}$. The second inequality is obtained by the relation
\begin{align}
\int \d \bs \beta\,p(\bs \beta ) \Tr \left(C_{\bs \beta}\bs \eta_{\omega}\bs \eta_{\omega}^{\rm T} C_{\bs \beta}+S_{\bs \beta}\bs \eta_{\omega}\bs \eta_{\omega}^{\rm T} S_{\bs \beta}\right)=& \int \d \bs \beta\,p(\bs \beta )\sum_{j=0}^{m-1}4|\beta_j|^2\omega_j\eta_j\\
\le& 4\max_j \omega_j\eta_j\sum_{j=0}^{m-1}\int \d \bs \beta\,p(\bs \beta )|\beta_j|^2\\
\le& 4m N_{\rm S}\max_j \omega_j\eta_j. 
\end{align}

Thereby, the rWMSE for an arbitrary classical strategy can be lower bounded as: 
\begin{align}
\epsilon_c &= \sqrt{\frac{\Tr[F_{\bs \theta,{\rm c}}^{-1}]}{m}}\\
&\ge \min_{\{p_i,|\phi_i\>\}}\sqrt{\frac{2N_{\rm B}+1}{4m^2 N_{\rm S}\max_j \omega_j\eta_j}\Tr\left[\left(\sum_{i=0}^{m-1} p_i |\phi_i\>\<\phi_i|\right)^{-1}\right]} \label{c11}\\
&= \min_{\{p_i,|\phi_i\>\}}\sqrt{\frac{2N_{\rm B}+1}{4m^2 N_{\rm S}\max_j \omega_j\eta_j}\sum_{i=0}^{m-1} p_i^{-1} } \\
&= \sqrt{\frac{2N_{\rm B}+1}{4 N_{\rm S}\max_j \omega_j\eta_j} },\label{c13}
\end{align}
where $\{p_i |\phi_i\>\}$ refers to an eigendecomposition for an arbitrary density matrix, the inequality is derived by the relation $\Tr[WB^{-1}]\ge \Tr[WA^{-1}],\,\,\forall A\ge B{\,\rm and\,} W\ge 0$. This proves Theorem~\ref{theo1}.

\subsection{Achievability of the lower bound}

Based on the inequality shown in Eq. (\ref{c11}), it can be inferred that the optimal QFIM exhibits a spectrum decomposition characterized by a uniform distribution.  However, this requirement cannot be satisfied if we also want to establish equality in the inequality shown in Eq. (\ref{c6}), as it achieves equality when applied to pure input states and the associated QFIM are rank-two matrices. 

To resolve this issue, one can consider a scenario where the experimenter implements $\nu$ rounds of experiments. In this case, we have the following bounds of QFIM: 
\begin{align}
F_{\bs \theta,{\rm c},\nu}&\le \frac {4mN_{\rm S}\max_j \omega_j\eta_j} {2N_{\rm B}+1} \nu \sum_{i=0}^{m-1} q_i |\psi_i\>\<\psi_i| , 
\end{align}
where $\{q_i,|\psi_i\>\}$ represents the eigen-decomposition for the density matrix associated to the summation of $\nu$ different QFIMs. In this case, one can achieve the lower bound of Eq. (\ref{c13}) without requiring full-rank QFIM for each round of the experiment. 

In the case with $\nu\ge \lceil m/2\rceil$, one can concentrate the displacement energy into two specific input pure modes in principle, in order to  achieve linear independence of the QFIMs: 
\begin{align}
F_{\bs \theta,{\rm c},\nu}&= \sum_{\{j,k\}} F_{{\rm c}}^{(j,k)},
\end{align}
where $F_{{\rm c}}^{(j,k)}$ is a rank-two QFIM that can be obtained by $\{\beta_j=\beta_ke^{-i(\theta_j-\theta_k+\text{arg}\omega_j/2-\text{arg}\omega_k/2+\pi/2)}=\sqrt{m N_{\rm S}/2},\beta_i=0,\forall i\neq j, k\neq j\}$. Then, each QFIM is bounded as follows: 
\begin{align}
F_{{\rm c}}^{(j,k)}
&= \frac {2mN_{\rm S} } {2N_{\rm B}+1} \left(\begin{matrix}
\omega_j\eta_j &0\\
0&\omega_k\eta_k
\end{matrix}\right).\label{c16}
\end{align}

In the ideal case $\nu= l m/2,\,\,\forall l\in \N$ and $m$ is even, we have:
\begin{align}
\left.\epsilon_c\right|_{\nu= l m/2,\,\,\forall l\in \N\, {\rm\ and\ m\ is \ even}} = & \sqrt{\frac{2N_{\rm B}+1}{4m\nu N_{\rm S}}\sum_{j=0}^{m-1} (\omega_j\eta_j)^{-1}}. 
\end{align}

In practice, if the values $\{\theta_j\}$ are completely unknown, one can alternatively devise more than $m$ experiments to avoid the design of two-mode experiments with fine-tuned phases. 

For the estimation of the average phase $\overline \theta':=\sum_{j=0}^{m-1} \sqrt{\omega_j\eta_j}\theta_j$, by applying the Jacobi transformation on the optimal QFIM of two parameters in Eq. (\ref{c16}), we will have the Fisher information as follows: 
\begin{align}
F_{\overline \theta',{\rm c}}&\le  \frac {4m\nu N_{\rm S}} {2N_{\rm B}+1}.
\end{align}
The corresponding RMSE is: 
\begin{align}
\epsilon_{\overline \theta',{\rm c}}&\ge  \sqrt{\frac{2N_{\rm B}+1} {4m\nu N_{\rm S}}}.
\end{align}

\subsection{Non-asymptotic parameter estimation with $\nu< \lceil m/2\rceil $}\label{app:non-asymptotic}

In practice, it is usually challenging to conduct more than $\nu\sim\lceil m/2\rceil$ experiments when the effective transmitter number $m$ is large. In this situation, one has to prepare a mixture of $|\map P|:=\lceil m/(2\nu)\rceil$ pure coherent states as the input to guarantee a full-rank QFIM. However, the upper bound in Eq. (\ref{c6}) will become loose with mixed probes due to the convexity of QFIM. Moreover, the QFIM is not applicable for evaluating errors in non-asymptotic cases. In this subsection, we will demonstrate a lower bound of error based on the Bayesian approach and an upper bound from homodyne detection.  

\subsubsection{Lower bound of error with Bayesian approaches}\label{app:non-asymptotic}

Given the fact that if the input is a $2m$-mode pure coherent state, the output state of the illumination network can only be encoded with two independent real parameters, the rWMSE for estimating each phase with single-shot measurement can be bounded as follows: 
\begin{align}
&\ \ \ \ \nonumber \left. \epsilon_{\rm c, single-shot} \right|_{\nu<\frac m 2 }\\&=\min_{\{M_a,\hat \theta_j(a),\map P\}} \sqrt{\frac {1}{m }\int \frac{\d\bs\theta}{(2\pi)^m}\sum_{\map P}\sum_{j\in\map P} \sum_{a}\Tr\left[\left(\sum_{h\in \map P}  \,p_{\bs \beta_h (j)}\, \rho_{\bs \beta_h}\right)\,M_{ a}\,\right]\left[\hat{\theta}_j(a)-{ \theta}_j\right]^2}\\
&\ge   \min_{\{M_a,\hat \theta_j^{(h)}(a),\hat \theta_{j,{\rm guess}},\map P\}} \sqrt{\begin{matrix}
\frac{1}{m} \int \frac{\d\bs\theta}{(2\pi)^m}\,\sum_{\map P}\sum_{j\in\map P} p_{\bs \beta_h(j)}\, \sum_{a}\Tr\left[\rho_{\bs \beta_h(j)}\,M_{ a}^{(h)}\,\right]\left[\hat{\theta}_j^{(h)}(a)-{ \theta}_j\right]^2\\
+\frac{1}{m} \int \frac{\d\bs\theta}{(2\pi)^m}\,\sum_{\map P}\sum_{ j\in\map P}\left[1-p_{\bs \beta_h(j)}\right]\,  \sum_{a}\Tr\left[\rho_{\bs \beta_h(j)}\,M_{ a}^{(h)}\,\right]\left[\hat{\theta}_{j,{\rm guess}}^{(h)}-{ \theta}_j\right]^2
\end{matrix}}\\
&\ge \min_{|\map P|,\{p_{\bs \beta_h(j)}\}}\sqrt{\frac{1}{m} \,\sum_{\map P} \sum_{j\in\map P}\left[p_{\bs \beta_h(j)}E_{\rm Baye}(\rho_{\bs\beta_h(j)},\theta_j)+\left(1-p_{\bs \beta_h(j)}\right) \frac{\pi^2}{3}\right]}\\
&= \sqrt{\frac{2\nu}{m}\min_j E_{\rm Baye}(\rho_{\bs\beta_h(j)},\theta_j) +\left(1-\frac{2\nu}{m}\right) \frac{\pi^2}{3}},
\end{align}
where $\bs\beta_h(j)$ is associated to the state that encodes the information about $\theta_j$ and $\map P$ denotes a mutually orthogonal set of $|\map P|\sim \frac{m}{2\nu }$ subscripts from 1 to $m$, $\rho_{\bs \beta_h}=D\left(\sum_{k=0}^{m-1}\sqrt{\omega_{k}\eta_{k}}e^{-i\theta_{k}}\beta_{k}^{(h)}\right)\rho_{N_{\rm B}}D\left(\sum_{l=0}^{m-1}\sqrt{\omega_{l}\eta_{l}}e^{-i\theta_{l}}\right.$ $\left.\beta_{l}^{(h)}\right)\otimes_{l=m}^{2m-1}|\beta_l^{(h)}\>\<\beta_l^{(h)}|$ is the output state of a classical illumination network when the input is a $2m$-mode pure coherent state $\otimes_{l=0}^{2m-1}|\beta_l\>$. Here, the state $\rho_{\bs \beta_h}$ carries information for two independent phases, while its mixture $\sum_h p_{\bs \beta_h}\rho_{\bs \beta_h}$ can encode $2|P|$ independent phases. $\hat \theta_j^{(h)}(a)$ refers to the unbiased estimator when the output is $\rho_{\bs\beta_h(j)}$. %Here, the first inequality will become tighter as the total energy increases $mN_{\rm S}\gg1$ and the background noise decreases $N_{\rm B}\ll 1$. In this case, the output state will be more distinguishable, thus increasing the success probability of establishing an unbiased estimator for each state $\rho_{\bs\beta_h(j)}$. 
Here, the second equation is derived from the fact that the state $\rho_{\bs\beta_h(j)}$ can only carry information for at most two independent real parameters. 
The inequality is derived by considering a uniform distribution of the phases and by using the relation $\min_{x\in [0,2\pi)}\int_{0}^{2\pi} \frac{\d \theta_j}{2\pi} (x-\theta_j)^2= \pi^2/3$ and the Bayesian bound for non-asymptotic measurement  \cite{helstrom1969quantum,personick1971application,yuen1973multiple,rubio2020bayesian}: 
$E_{\rm Baye}(\rho(\bs\theta),\theta_j) =\int \d \bs \theta\, p(\bs \theta)\theta_j^2-  \Tr\left[\rho(\bs\theta)\, B_j^2\right] $ where $\{B_j\}$ are Bayesian counterpart of the equation for the SLD defined by: 
$B_j \int \d \bs \theta\, p(\bs \theta)\rho (\bs \theta )+ \int \d \bs \theta\,  p(\bs \theta)\rho (\bs \theta )\, B_j=2\int \d \bs \theta\,  p(\bs \theta)\rho (\bs \theta )\theta_j$ with $p(\bs\theta )$ being the uniform distribution. This Bayesian bound can be evaluated numerically by solving the above Sylvester equation. 

More generally, if we choose to estimate $J=\lfloor 2\nu/\nu' \rfloor$ phases and leave the other for random guess, we have the error: 
\begin{align}
\left. \epsilon_{\rm c, multiple-shot} \right|_{\nu<\frac m 2 }&=\min_J \sqrt{\frac{J}{m}\frac{\min_j E_{\rm Baye}(\rho_{\bs\beta_h(j)},\theta_j)}{\nu'} +\left(1-\frac{J}{m}\right) \frac{\pi^2}{3}},
\end{align}
where $E_{\rm Baye}(\rho_{\bs\beta_h(j)},\theta_j)$ is the single-shot MSE of estimating $j$ for the probe with displacement $\sqrt{\omega_k\eta_kN_{\rm S}/2}e^{-i\theta_k}+\sqrt{\omega_j\eta_jN_{\rm S}/2}e^{-i\theta_j}$ for arbitrary $j\neq k$. 

\subsubsection{Upper bound of error with pure probe and homodyne measurement}

Consider the situation where the experimenter prepares a highly-bright single-mode probe for the estimation of a single phase. The returning state will be: 
\begin{align}
\rho_{k}=D\left(\,e^{-i\theta_{k}}\sqrt{ {m\,\omega_{k}\eta_{k}N_{\rm S}}  }\,\right)\rho_{N_{\rm B}}D^\dag \left(\,e^{-i\theta_{k}}\sqrt{ {m\,\omega_{k}\eta_{k}N_{\rm S}}  }\,\right). 
\end{align}
Then, the experimenter performs a single-shot homodyne measurement, which produces the probability distribution  \cite{serafini2017quantum}: 
\begin{align}\label{eqc15}
p( q )
&=\frac{\exp\left[-\frac 1 {2(2N_{\rm B}+1)} \left( q -\cos\theta_{k}\sqrt{ {m\,\omega_{k}\eta_{k}N_{\rm S}}  }\,\right)^{2}  \right]}{ \sqrt{2\pi \left(2N_{\rm B}+1\right)}},
\end{align}
where $q\in \R$ refers to the homodyne  measurement result. 

By constructing the maximum likelihood estimator: 
\begin{align}
\hat \theta_k  (q)=
\begin{cases}
\text{arccos}\left(q\,(m\,\omega_k\eta_k N_{\rm S})^{-1/2}\right),& \frac{|q|}{\sqrt{m\,\omega_k\eta_k N_{\rm S}}}\le 1\\
\frac{3\pi}{2},& \frac{|q|}{\sqrt{m\,\omega_k\eta_k N_{\rm S}}}> 1
\end{cases}
\end{align}
the mean-square-error will be: 
\begin{align}
E_{\rm c,homo,k}&=\int_{0}^{2\pi}  \frac{\d \theta_k}{2\pi} \int_{-\infty}^\infty \d q\, p (q) \left[\hat \theta_k  (q)-\theta_k \right]^2\\
&=
\begin{matrix}
\int_{0}^{2\pi}  \frac{\d \theta_k}{2\pi} \int_{\frac{|q|}{\sqrt{m\,\omega_k\eta_k N_{\rm S}}}\le 1} \d q\, \frac{\exp\left[-\frac 1 {2(2N_{\rm B}+1)} \left( q -\cos\theta_{k}\sqrt{ {m\,\omega_{k}\eta_{k}N_{\rm S}}  }\,\right)^{2}  \right]}{ \sqrt{2\pi \left(2N_{\rm B}+1\right)}}\left[\text{arccos}\left(q\,(m\,\omega_k\eta_k N_{\rm S})^{-1/2}\right)-\theta_k \right]^2\\
+\int_{0}^{2\pi}  \frac{\d \theta_k}{2\pi} \int_{\frac{|q|}{\sqrt{m\,\omega_k\eta_k N_{\rm S}}}> 1} \d q\, \frac{\exp\left[-\frac 1 {2(2N_{\rm B}+1)} \left( q -\cos\theta_{k}\sqrt{ {m\,\omega_{k}\eta_{k}N_{\rm S}}  }\,\right)^{2}  \right]}{ \sqrt{2\pi \left(2N_{\rm B}+1\right)}}\left(\frac{3\pi}{2}-\theta_k \right)^2
\end{matrix}\\
&=
\begin{matrix}
c\int_{0}^{2\pi}  \frac{\d \theta_k}{2\pi} \int_{|q|\le 1} \d q\, \frac{\exp\left[-\frac 1 {2(2N_{\rm B}+1)} \left( q -\cos\theta_{k}\,\right)^{2} c^2  \right]}{ \sqrt{2\pi \left(2N_{\rm B}+1\right)}}\left[\text{arccos}\left(q\right)-\theta_k \right]^2\\
+c\int_{0}^{2\pi}  \frac{\d \theta_k}{2\pi} \int_{|q|> 1} \d q\, \frac{\exp\left[-\frac 1 {2(2N_{\rm B}+1)} \left( q -\cos\theta_{k}\,\right)^{2}c^2  \right]}{ \sqrt{2\pi \left(2N_{\rm B}+1\right)}}\left(\frac{3\pi}{2}-\theta_k \right)^2
\end{matrix}\\
&=
\begin{matrix}
c\int_{0}^{2\pi}  \frac{\d \theta_k}{2\pi} \int_{0}^{\pi} \d \theta_k'\,\sin \theta_k' \frac{\exp\left[-\frac 1 {2(2N_{\rm B}+1)} \left( \cos \theta_k' -\cos\theta_{k}\,\right)^{2} c^2  \right]}{ \sqrt{2\pi \left(2N_{\rm B}+1\right)}}\left(\theta_k'-\theta_k \right)^2\\
+\int_0^{2\pi} \frac{\d \theta_k}{2\pi}\frac{1}{2} \left[2-\text{erf}\left(\frac{c(1-\cos \theta_k)}{\sqrt{2(2N_{\rm B}+1)}}\right)-\text{erf}\left(\frac{c(1+\cos \theta_k)}{\sqrt{2(2N_{\rm B}+1)}}\right)\right]\left(\frac{3\pi}{2}-\theta_k\right)^2\end{matrix},
\end{align}
where $c=\sqrt{m\,\omega_k\eta_k N_{\rm S}}$ is known. Thereby, by repeating experiment $\nu'$ times to estimate a single parameter, the mean-square-error will decrease in a scaling $\mathcal O(1/\sqrt{\nu'})$. 

Consider the non-asymptotic multi-parameter sensing case where the experimenter only estimate $J'$ parameters, for each implementing $\nu'$ rounds of experiments. Then, he will leave  the other independent parameter at random guess. Therefore, the overall rWMSE will be: 
\begin{align}
\left.\epsilon_{\rm c,homo}\right|_{\nu<m}&=\min_{J'}\sqrt{\frac 1 m \sum_{k=0}^{J'-1} \frac{E_{\rm c,homo,k}}{\nu'}+\left(1-\frac {J'} m \right)\frac{\pi^2}{3}},
\end{align}
where $\nu=\nu'J'< m $ is the total experiment rounds.

\section{Quantum illumination network set-up and its measurement design}\label{APP: QI-setup}

\subsection{Representation of Gaussian quantum systems}
In this section, we review basic definitions for continuous-variable quantum systems, showing the Gaussian representation of quantum states in terms of their moments. 

Consider a $2m$-mode continuous-variable Gaussian quantum system, where the whole system can be fully described by its first and second moments \cite{weedbrook2012gaussian,serafini2017quantum}. Specifically, if we adopt the natural units $\hbar=2$, the characteristic function of a  $2m$-mode Gaussian state $\rho\in\textbf{St}(\map H^{\otimes 2m})$ can be defined as: 
\begin{align}
\chi (\bs{\xi}')&=\Tr[\rho D(\bs{\xi}')]\\
&=\exp\left[-\frac 1 2 \bs{\xi}'^{\rm T}\bs{\Omega}^{\rm T}\bs{V}\bs{\Omega}\bs{\xi}'+i(\bs{\Omega}\bs{\xi}')^{\rm T}\bs{ \xi}\right],
\end{align}
where the matrix $\bs{ \Omega}=\left(\begin{matrix}
0&-1\\1&0
\end{matrix}\right)\otimes\mathbb I_{2m}$ is known as the symplectic form, $D(\bs{\xi}')$ us the displacement operator, $\bm V$ is the covariance matrix and $\bm xi$ is the mean.
The displacement operator $D(\bs{\xi}')=\exp(i\bs{ r}^{\rm T}\bs{\Omega}\bs{\xi}')$, where $\bs{\xi}'\in \R^{4m}$ are quadratures to describe quasidistributions and $\bs{ r}=({\hat q}_{0},{\hat p}_0,\cdots,{\hat q}_{2m-1},{\hat p}_{2m-1})^{\rm T}$ refers to the quadrature field operator. 
The covariance matrix defined by $V_{ij}= \Tr[\{\Delta \hat r_{i},\Delta\hat r_j\}\rho]/2, (i,j=0,\cdots,2m-1)$ with $\Delta \hat r_{i}=\hat r_{i}-\bs \xi_i$ and $\{A,B\}=AB+BA$ being the anticommutator, while the mean $\bs \xi=\Tr[\bs r\rho]$.

\subsection{Quantum illumination network}\label{APP:setup_CtoD protocol}

The quantum illumination network that involves the CtoD measurement protocol consists of the following steps: 
\begin{enumerate}
\item  [(I)] \emph{Preparation of  probes:---} A transmitter array   generates $m$ copies of  two-mode squeezed vacuum (TMSV) states. The overall state can be described by the  displacement and covariance matrix \cite{weedbrook2012gaussian}: 
\begin{align}
\begin{cases}
\bs \xi_{\rm I}&=\bs 0,\\
{\bs V}_{\rm I}&=\mathbb I_m \otimes  \left(\begin{matrix}
(2N_{\rm S}+1)\mathbb I_2 &2\sqrt{N_{\rm S}(N_{\rm S}+1)}\mathbb Z\\
2\sqrt{N_{\rm S}(N_{\rm S}+1)}\mathbb Z&(2N_{\rm S}+1)\mathbb I_2
\end{matrix}\right),
\end{cases}
\end{align}
where $\bs 0=(0,\cdots,0)^{\rm T}$ is the vector with zeros, $\mathbb Z=\left(\begin{matrix}
1 &0\\0&-1
\end{matrix}\right)$ is the Pauli-Z matrix, $\mathbb I_2$ is the $2\times 2$ identity, $\mathbb I_N$ is the $N\times N$ identity, and $N_{\rm S}$ refers to the average photon number of the two-mode squeezed vacuum state.

\item [(II)] \emph{Sensing of phase and reflectivity:---} The signal modes of the entangled states are sent to the target, and each independently experiences a thermal loss channel as described by Eq. (\ref{thermal-loss}) in the main text. The overall $m$-mode noisy process gives rise to the following moments:  
\begin{align}
\begin{cases}
\bs \xi_{\rm II}&=\bs 0,  \\
\bs V_{\rm II}&=\bigoplus_{j=0}^{m-1} \left(\begin{matrix}
(2\eta_j  N_{\rm S}+2N_{\rm B}+1)\mathbb I_2 &2\sqrt{\eta_j N_{\rm S}(N_{\rm S}+1)}\mathbb R_j  \mathbb Z\\
2\sqrt{\eta_j  N_{\rm S}(N_{\rm S}+1)}\mathbb Z\mathbb R_j ^{\rm T}&(2N_{\rm S}+1)\mathbb I_2
\end{matrix}\right),
\end{cases}
\end{align}
with $\mathbb R_j $ being an operator defined as $\mathbb R_j =\cos \theta_j  \mathbb I_2 -i \sin \theta_j  \mathbb Y$, and $\mathbb Y=\left(\begin{matrix}
0&-i\\i&0
\end{matrix}\right)$ being the Pauli-Y operator. 
\item [(III)] \emph{Coherent interference:---} The interference among multiple returning modes can be modeled by a linear operation wherein the $m$ signal-imprinted modes are taken as input. This channel will lead to the following moments: 
\begin{align}
\begin{cases}
\bs \xi_{\rm II}=&\bs 0,  \\
\bs V_{\rm III}=& \frac 1 2 P_m(W\otimes \mathbb I_2\oplus  \mathbb I_m\otimes \mathbb I_2)\left(\mathbb I_{m_{\rm re}-m}\otimes \mathbb I_2\oplus\bs V_{\rm II}\right) (W^\dag\otimes \mathbb I_2\oplus   \mathbb I_m\otimes \mathbb I_2)P_m\\
&+\frac 1 2 P_m\left[(W\otimes \mathbb I_2\oplus  \mathbb I_m\otimes \mathbb I_2)\left(\mathbb I_{m_{\rm re}-m}\otimes \mathbb I_2\oplus\bs V_{\rm II} \right) (W^\dag\otimes \mathbb I_2\oplus   \mathbb I_m\otimes \mathbb I_2)\right]^{\rm T}P_m,
\end{cases}
\end{align}
where $W$ is a $m_{\rm re}\times m_{\rm re}$ unitary matrix with $\sqrt{\omega_{ij}}\in \R$ being the entry at its $i$-th row and $j$-th column, satisfying the condition $\sum_{n=0}^{m_{\rm re}-1} \sqrt{\omega_{in}}\sqrt{\omega_{kn}}=\delta_{ik},\forall i,n,k=0,\cdots,m_{\rm re}-1$, $P_m$ is the projector to the $m$ signaling modes and the $m$ idlers. For algebraic convenience, we assume the condition $\omega_{ij}\sim 1/\sqrt{m_{\rm re}}$ when estimating error scalings. 
\item [(IV)] \emph{Discarding of m-1 modes:---} Due to the multi-access channel, the experimenter has only access to one of the modes after interference and therefore equivalently discards the $(m-1)$ imprinted modes and obtains a state with the following moments: 
\begin{align}\label{eqb6}
\begin{cases}
\bs \xi_{\rm IV} &= \bs 0  \\
\bs V_{\rm IV}\\
=&\left(\begin{matrix}
\left(2\sum_{j=0}^{m-1}\omega_{j,m-1}\left(\eta_{j} N_{\rm S}+N_{\rm B}\right)+1\right)\mathbb I_2 &S_{0}^{\rm T}&\cdots &S_{m-1}^{\rm T}\\
S_{0} & (2N_{\rm S}+1)\mathbb I_2 & \cdots &0\\
\vdots  &\vdots& \ddots   &\vdots \\
S_{m-1}& 0& \cdots  &(2N_{\rm S}+1)\mathbb I_2 \\
\end{matrix}\right),
\end{cases}
\end{align}
where the matrix entries are defined as: $S_{j}= 2\sqrt{\omega_{j,m-1}\eta_{j} N_{\rm S}(N_{\rm S}+1)}\mathbb Z\mathbb R_{j}^{\rm T} $. 
\end{enumerate}

In the following part of the Appendix and the main text, for the simplicity of notations, we denote the interference coefficients $\omega_{j,m-1}$ by $\omega_j$ and assume that they are all positive numbers.

\subsection{Parametric-amplifier (PA) receiver network}\label{APP: OPA}

Let us examine two practical methods for implementing the PA receiver network, known as the parallel (pPCR) and serial phase-conjugate receiver (sPCR), as shown in Figure~\ref{fig:OPA_PCR(app)}. %Without losing the generality, let's first consider single-shot measurement design and then extend it to multiple-copy scenarios. 

\subsubsection{Parallel PCR}

The parallel PCR scheme is illustrated in Figure \ref{fig:OPA_PCR(app)} (a). Given the output state from the QI network (as shown in Eq. (\ref{eqb6})), the pPCR consists of the following steps: 
\begin{enumerate}
\item [(V-]A1) {\em PA on the returning mode and a vacuum state}:--- The experimenter conducts a joint PA operation that generates interference between the returning mode and a vacuum state: 
\begin{align}
\begin{cases}
\hat q_{\rm R}\to \sqrt g \, \hat q_{\rm vac}+ \sqrt{g-1}\, \hat q_{\rm R}\\
\hat p_{\rm R}\to \sqrt g\, \hat p_{\rm vac}- \sqrt{g-1}\, \hat p_{\rm R}\\
\hat q_{\rm vac}\to \sqrt {g-1} \, \hat q_{\rm vac}+ \sqrt{g}\, \hat q_{\rm R}\\
\hat p_{\rm vac}\to -\sqrt {g-1}\, \hat p_{\rm vac}+ \sqrt{g}\, \hat p_{\rm R},\\
\end{cases}\label{eqb7}
\end{align}
where $q_{\rm R}(p_{\rm R})$ refers to the position (momentum) operator of the returning mode, $q_{\rm vac}(p_{\rm vac})$ refers to the position (momentum) operator of the vacuum whose covariance matrix is $\mathbb I_2$, and $g\ge 1$ is the amplification gain.  After discarding one mode of PA, the covariance matrix of the $(m+1)$-mode state will be transformed into the following matrix: 
\begin{align}
\bs V_{\rm V-A1 }= &\left(\begin{matrix}
\left[g+(g-1)\left(2\sum_{j=0}^{m-1}\omega_j(\eta_{j} N_{\rm S}+N_{\rm B})+1\right)\right]\mathbb I_2 &\sqrt{g-1}\,\mathbb Z S_{0}^{\rm T}&\cdots &\sqrt{g-1}\,\mathbb Z S_{m-1}^{\rm T}\\
\sqrt{g-1}\,S_{0} \mathbb Z& (2N_{\rm S}+1)\mathbb I_2 & \cdots &0\\
\vdots  &\vdots& \ddots   &\vdots \\
\sqrt{g-1}\,S_{m-1}\mathbb Z& 0& \cdots  &(2N_{\rm S}+1)\mathbb I_2 \\
\end{matrix}\right).\label{eqb8}
\end{align}
\item [(VI-]A1) {\em Distribution via a multi-port beamsplitter}:--- By using a multi-port beamsplitter, one can obtain the following %transformation:
%\begin{align}\begin{cases}\hat q_{\rm R}\to \hat q_{\rm R_j}:=\left(c_{j,m-1}\hat q_{\rm R}+\sum_{k=0}^{m-2} c_{jk}\,\hat q_{{\rm vac}_k} \right)\\\hat p_{\rm R}\to \hat p_{\rm R_j}:=\left(c_{j,m-1}\hat p_{\rm R}+\sum_{k=0}^{m-2} c_{jk}\,\hat p_{{\rm vac}_k} \right)\end{cases},\ \ \left(j=0,\cdots,m-1;\, \sum_k |c_{jk}|^2=\frac{m-1}{m}\right)\end{align}which leads to the 
covariance matrix: 
\begin{align}
\bs V_{\rm VI-A1 }= &\frac 1 2 (U\otimes \mathbb I_2\oplus  \mathbb I_m\otimes \mathbb I_2)\left(\mathbb I_{m-1}\oplus \bs V_{\rm VI.A1 }\right)(U^\dag\otimes \mathbb I_2\oplus   \mathbb I_m\otimes \mathbb I_2)\\
&+\frac 1 2 \left[(U\otimes \mathbb I_2\oplus  \mathbb I_m\otimes \mathbb I_2)\left(\mathbb I_{m-1}\oplus \bs V_{\rm VI.A1 }\right) (U^\dag\otimes \mathbb I_2\oplus   \mathbb I_m\otimes \mathbb I_2)\right]^{\rm T}, 
\end{align}
where $U$ is a $m\times m$ unitary matrix with $\sqrt{u_{ij}}\in\R$ being the entry at its $i$-th row and $j$-th column, satisfying the condition $\sum_{n=0}^{m-1} \sqrt{u_{in}}\sqrt{u_{mn}}=\delta_{im},\forall i,m=0,\cdots,m-1$. Without losing the generality, let's consider the multi-port beamsplitter with $|u_{m-1,j}|=1/m,\forall j=0,\cdots,m-1$. Then, the covariance matrix becomes: 
\begin{align}
\bs V_{\rm VI-A1 }  =&\left(\begin{matrix}
\left(\frac{2N_{{\rm B},g}'}m+1\right)\mathbb I_2&\frac{2N_{{\rm B},g}'}{\sqrt{m}}\mathbb I_2&\cdots &\frac{2N_{{\rm B},g}'}{\sqrt{m}}\mathbb I_2&\sqrt{\frac{g-1}m}\,\mathbb Z S_{0}^{\rm T}&\cdots &\sqrt{\frac{g-1}m}\,\mathbb Z S_{m-1}^{\rm T}\\
\frac{2N_{{\rm B},g}'}{\sqrt{m}}\mathbb I_2&\ddots &&\vdots&\vdots&&\vdots\\
\vdots &&\ddots &\frac{2N_{{\rm B},g}'}{\sqrt{m}}\mathbb I_2 &&&\\
\frac{2N_{{\rm B},g}'}{\sqrt{m}}\mathbb I_2 &\cdots  &\frac{2N_{{\rm B},g}'}{\sqrt{m}}\mathbb I_2&\left(\frac{2N_{{\rm B},g}'}m+1\right)\mathbb I_2 &\sqrt{\frac{g-1}m}\,\mathbb Z S_{0}^{\rm T}&\cdots &\sqrt{\frac{g-1}m}\,\mathbb Z S_{m-1}^{\rm T}\\
\sqrt{\frac{g-1}m}\,S_{0} \mathbb Z&\cdots&&\sqrt{\frac{g-1}m}\,S_{0} \mathbb Z& (2N_{\rm S}+1)\mathbb I_2 & \cdots &0\\
&&&\vdots  &\vdots& \ddots   &\vdots \\
\sqrt{\frac{g-1}m}\,S_{m-1}\mathbb Z&\cdots&&\sqrt{\frac{g-1}m}\,S_{m-1}\mathbb Z& 0& \cdots  &(2N_{\rm S}+1)\mathbb I_2 \\
\end{matrix}\right),
\end{align}
where $N_{{\rm B},g}'=(g-1)(N_{\rm B}'+1)$ and $N_{\rm B}'=\sum_{j=0}^{m-1}\omega_j(\eta_{j} N_{\rm S}+N_{\rm B})$. 
\item [(VII]-A1) {\em Interfere the signal modes and idlers pairwisely with balanced beamsplitters}:--- Implement balanced beamsplitter operations: $a\to (a+b)/\sqrt 2, b\to (a-b)/\sqrt 2$ pairwisely on each output from the multi-port beamsplitter and one of the idler modes. Output states of the $j$-th balanced beamsplitter can be denoted by the subscript $(j,\pm)$. The resulting covariance matrix is denoted by $\bs V_{\rm VII-A1 }$. 

Now let's define $\hat N_{j,\pm}:=a^\dag_{j,\pm}a_{j,\pm}$ as the photon number operator of the ($j,\pm$) mode. Thanks to the results in Refs.~\cite{dodonov1994multidimensional,parthasarathy2015particle,vallone2019means}, we can express the first and second moments of the photon number operator $\{\hat N_{j,\pm}\}$ as functions of the covariance matrix $\bs V_{\rm VII-A1 }$. Further, if we define the photon number difference operator at the $j$-th beamsplitter as $\hat N_{j,{\rm D}}:=\hat N_{j,+}-\hat N_{j,-}$, we will have the following moments of $\{\hat N_{j,{\rm D}}\}$:  
\begin{align}
\left\<\hat N_{j,{\rm D}}\right>=&\left\<\hat N_{j,{\rm +}}\right>-\left\<\hat N_{k,{\rm -}}\right>= 2\sqrt{\frac{(g-1)\omega_j\eta_j N_{\rm S}(N_{\rm S}+1)}m}\cos \theta_j , \label{eqb13}\\
\left.\left\<\hat N_{j,{\rm D}}\hat N_{k,{\rm D}}\right>-\left\<\hat N_{j,{\rm D}}\right>\left\<\hat N_{k,{\rm D}}\right>\right|_{j\neq k}= &\left\<\hat N_{j,{\rm +}}\hat N_{k,{\rm +}}\right>-\left\<\hat N_{j,{\rm -}}\hat N_{k,{\rm +}}\right>-\left\<\hat N_{j,{\rm +}}\hat N_{k,{\rm -}}\right>+\left\<\hat N_{j,{\rm -}}\hat N_{k,{\rm -}}\right>\nonumber \\
&-\left\<\hat N_{j,{\rm +}}\right\>\left\<\hat N_{k,{\rm +}}\right>+\left\<\hat N_{j,{\rm -}}\right\>\left\<\hat N_{k,{\rm +}}\right>+\left\<\hat N_{j,{\rm +}}\right\>\left\<\hat N_{k,{\rm -}}\right>-\left\<\hat N_{j,{\rm -}}\right\>\left\<\hat N_{k,{\rm -}}\right>\\
=&\frac{2(g-1)}{m} N_{\rm S}(N_{\rm S}+1)\sqrt{\omega_j\omega_k\eta_j\eta_k}\cos\theta_j\cos\theta_k,\\
\left\<\hat N_{j,{\rm D}}\hat N_{j,{\rm D}}\right>-\left\<\hat N_{j,{\rm D}}\right>\left\<\hat N_{j,{\rm D}}\right>=&\frac {g-1} m (N_{\rm B}'+1)(2N_{\rm S}+1) +N_{\rm S} +\frac {2(g-1)} m  N_{\rm S}(N_{\rm S}+1)\omega_j\eta_j\cos^2\theta_j, \label{eqb16}
\end{align}

\iffalse 
{\color{red} The following are the original calculation. Leave it unchanged. }

The local states at the output ports are thermal states $\rho_{i,\pm}=1/(N_{j,\pm}+1)\sum_{n=0}^\infty [N_{j,\pm}/(N_{j,\pm}+1)]^n |n\>\<n|$ with  average photon numbers: 
\begin{align}
N_{j,\pm}&:=\left\<a_{j,\pm}^\dag a_{j,\pm}\right\>\\
&= \frac {g-1} {2m} \left(N_{\rm B}'+1\right) +\frac{N_{\rm S}} 2 \pm  \sqrt{\frac{(g-1)\omega_j\eta_j N_{\rm S}(N_{\rm S}+1)}m}\cos \theta_j \label{eqb11} ,\ \ \ j=0,\cdots,m-1 ,
\end{align}
where $N_{\rm B}'=N_{\rm B}+\sum_{j=0}^{m-1}\omega_j\eta_{j} N_{\rm S}$. In addition, the off-diagonal elements of resulting covariance matrix indicate the following cross-correlation: 
\begin{align}
N_{j,{\rm cor}}&:=
\frac 1 2 \left\<\left(a_{j,+}^\dag a_{j,-}+a_{j,-}^\dag a_{j,+}\right)\right\>\\
&=\frac {g-1} {2m} \left(N_{\rm B}'+1\right) -\frac{N_{\rm S}} 2 . \label{eqb14aa}
\end{align}

Moreover, we have the higher order moments: 
\begin{align}
\left\<\left(a^\dag_{j,\pm} a_{j,\pm}\right)^2\right\>&=N_{j,\pm}\left(N_{j,\pm}+1\right)\\
\left\<a^\dag_{j,+} a_{j,+} a^\dag_{j,-} a_{j,-}\right\>&\ge N^2_{j,{\rm cor}}
\end{align}
where the first equation is derived by computing the variance of photon number for thermal states, the second equation is obtained by applying the Cauchy–Schwarz inequality.  
\fi

\item [(IX]-A1) {\em Detecting the difference between the total photon 
counts }:--- Given $\nu\gg 1$ rounds of experiments, estimate physical parameters $\{\theta_j,\eta_j\}$ from the difference between total photon numbers from the $\{+,-\}$ output ports of each balanced beamsplitter. Given the multidimensional central limit theorem, the total photon number difference $ N_{j,{\rm D}}^{\rm tot}:= \sum_{n=0}^{\nu-1}\left\<\hat N_{j,{\rm D}}^{(n)}\right\>$ of the $j$-th beamsplitter follows the distribution: 
\begin{align}\label{eqb17}
p\left(\bs N_{{\rm D}}^{\rm tot}\right)&\sim\frac{1}{ \sqrt{(2\pi)^{ m  }\det \bs \Sigma}} \exp \left[-\frac 1 2 \left(\bs N_{{\rm D}}^{\rm tot}-\bs \mu\right)^T \bs \Sigma^{-1}\left(\bs N_{{\rm D}}^{\rm tot}-\bs \mu\right)\right]\\
\bs \mu& = \, \nu\cdot \left\<\bs{\hat N}_{{\rm D}} \right\>\\
\bs \Sigma&=\,\nu\cdot \left\<\left(\bs {\hat N}_{{\rm D}}^{\rm tot}-\bs \mu\right) \bs \left(\bs {\hat N}_{{\rm D}}^{\rm tot}-\bs \mu\right)^T\right\>,
\end{align}
where $\bs { N}_{{\rm D}}^{\rm tot}:=(  N_{0,{\rm D}}^{\rm tot},\cdots, N_{m-1,{\rm D}}^{\rm tot})^T$ is the vector of total photon count difference, $\bs {\hat N}_{{\rm D}}^{\rm tot}:=( \hat N_{0,{\rm D}}^{\rm tot},\cdots,\hat N_{m-1,{\rm D}}^{\rm tot})^T$ refers to the photon number difference operators of a single experiment, the explicit expressions of $\bs m$ and $\bs\Sigma$ are given by Eqs. \ref{eqb13} and \ref{eqb16}.

\end{enumerate}

\subsubsection{Serial PCR}

Consider the serial phase-conjugate receiver as illustrated in Figure \ref{fig:OPA_PCR(app)}. (b). It would consist the following steps: 
\begin{enumerate}
\item [(V]-A2) {\em Conduct a joint PA operation to interfere the returned state with a vacuum state}:--- The experimenter implements a joint PA operation that takes the first vacuum state and the returning mode as the input. The covariance matrix can be updated following Eq. (\ref{eqb7}): 
\begin{align}
\bs V_{\rm V-A2 }= &\left(\begin{matrix}
\left[g-1+g\left(2N_{\rm B}'+1\right)\right]\mathbb I_2&2\sqrt{g(g-1)}\left(N_{\rm B}'+1\right)\mathbb Z &\sqrt{g}\, S_{0}^{\rm T}&\cdots &\sqrt{g}\, S_{m-1}^{\rm T}\\2
\sqrt{g(g-1)}\left(N_{\rm B}'+1\right)\mathbb Z&\left[g+(g-1)\left(2N_{\rm B}'+1\right)\right]\mathbb I_2 &\sqrt{g-1}\,\mathbb Z S_{0}^{\rm T}&\cdots &\sqrt{g-1}\,\mathbb Z S_{m-1}^{\rm T}\\
\sqrt{g}\, S_{0}&\sqrt{g-1}\,S_{0} \mathbb Z& (2N_{\rm S}+1)\mathbb I_2 & \cdots &0\\
\vdots &\vdots  &\vdots& \ddots   &\vdots \\
\sqrt{g}\, S_{0}&\sqrt{g-1}\,S_{m-1}\mathbb Z& 0& \cdots  &(2N_{\rm S}+1)\mathbb I_2 \\
\end{matrix}\right).
\end{align}
\item [(VI-]A2) {\em Repeated PA}:--- Store one of the output states from the PA operation. Provide the other output state and an additional vacuum state to a subsequent PA (see Figure \ref{fig:OPA_PCR(app)}). Repeat the aforementioned step for $m$ times in total. Finally, the covariance matrix becomes:  
\begin{align}
&\nonumber \bs V_{\rm VI-A2 } \\ =&\left(\begin{matrix}
\left(2g^{m-1}N_{{\rm B},g}'+1\right)\mathbb I_2&\sqrt{g^{2m-3}}u&\cdots &\sqrt{g^{m-1}}u &\sqrt{g^{m-1}}\,v_{0}&\cdots &\sqrt{g^{m-1}}\,v_{m-1}\\
\sqrt{g^{2m-3}}u&\ddots &&\vdots&\vdots&&\vdots\\
\vdots &&\left({2gN_{{\rm B},g}'}+1\right)\mathbb I_2 &\sqrt{g}u &\sqrt{g}v_{0}&\cdots &\sqrt{g}v_{m-1}\\
\sqrt{g^{m-1}}u &\cdots  &\sqrt{g}u &\left({2N_{{\rm B},g}'}+1\right)\mathbb I_2 &v_0&\cdots &v_{m-1}\\
\sqrt{g^{m-1}}v_{0}^\dag&\cdots&\sqrt{g}v_{0}&v_{0}& (2N_{\rm S}+1)\mathbb I_2 & \cdots &0\\
&&\vdots&\vdots  &\vdots& \ddots   &\vdots \\
\sqrt{g^{m-1}}v_{m-1}^\dag&\cdots&\sqrt{g}v_{m-1}&v_{m-1}& 0& \cdots  &(2N_{\rm S}+1)\mathbb I_2 \\
\end{matrix}\right),
\end{align}
where $N_{{\rm B},g}'=(g-1)(N_{\rm B}'+1)$, $u=2N_{{\rm B},g}'\mathbb I_2$, and  $v_{j}=\sqrt{g-1}\mathbb Z S_j^{\rm T}$. 
\item [(VII-]A2) {\em Interfere the output states of PA and idlers pairwisely with balanced beamsplitters}:--- The experimenter implements balanced beamsplitter operations $a\to (a+b)/\sqrt 2, b\to (a-b)/\sqrt 2$ to pairwisely generate interference between the stored output states of PAs and the idler states. Output states of the $j$-th balanced beamsplitter can be denoted by the subscript
$(j, \pm)$. %The resulting covariance matrix is denoted by $\bs V_{\rm VII-A2}$. 

Then, the first and second moments of the photon number difference operator $\{\hat N_{j,{\rm D}}:=a_{j,+}^\dag a_{j,+}-a_{j,-}^\dag a_{j,-}\}$ are as follows: 
\begin{align}
\left\<\hat N_{j,{\rm D}}\right>=&2\sqrt{g^{j}(g-1)\omega_j\eta_jN_{\rm S}(N_{\rm S}+1)}\cos\theta_j,\\
\left.\left\<\hat N_{j,{\rm D}}\hat N_{k,{\rm D}}\right>-\left\<\hat N_{j,{\rm D}}\right>\left\<\hat N_{k,{\rm D}}\right>\right|_{j\neq k}=&2(g-1)N_{\rm S}(N_{\rm S}+1)\sqrt{g^{j+k}\omega_j\omega_k\eta_j\eta_k }\cos\theta_j\cos\theta_j,\\
\left\<\hat N_{j,{\rm D}}\hat N_{j,{\rm D}}\right>-\left\<\hat N_{j,{\rm D}}\right>\left\<\hat N_{j,{\rm D}}\right>=&g^{j}N_{{\rm B},g}'(2N_{\rm S}+1)+N_{\rm S}+2 g^{j}(g-1)N_{\rm S}(N_{\rm S}+1)\omega_j\eta_j\cos^2\theta_j.
\end{align}
\item [(VIII]-A1) {\em Detecting the difference between the total photon 
counts }:--- Given $\nu\gg 1$ rounds of experiments, estimate physical parameters $\{\theta_j,\eta_j\}$ from the difference between total photon numbers from the $\{+,-\}$ output ports of each balanced beamsplitter. Given the multidimensional central limit theorem, the probability of total photon number difference follows Eq. (\ref{eqb17}).  
\end{enumerate}

\begin{figure}[t]
\includegraphics[width=0.85\textwidth,trim=2 2 2 20,clip]{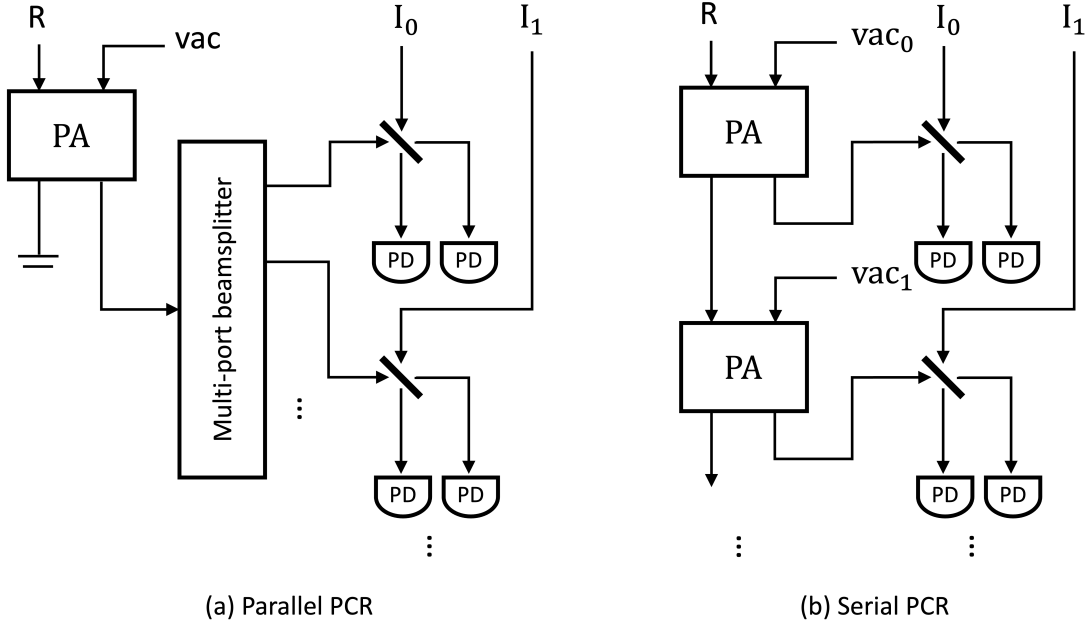}
\caption[]{{Schematic of two practical PA receiver designs. }(a) serial phase-conjugate receiver (sPCR), (b) parallel phase-conjugate receiver (pPCR). 
}\label{fig:OPA_PCR(app)}
\end{figure}

\subsection{Correlation-to-displacement (CtoD) conversion}

Let's look at the measurement design with correlation-to-displacement conversion. After obtaining the overall output state in Eq. (\ref{eqb6}), one can implement the following step: 
\begin{enumerate}
\item [(V.B)] \emph{Measuring one returning mode:---} The experimenter performs a \emph{single-mode heterodyne measurement} $\{|\chi \>\<\chi|| \chi =(q_x +i p_x)/2\in \C\}$ 
%for homodyne measurement see: PHYSICAL REVIEW A 81, 062338 
on the output of the multiple-access channel. For simplicity of notation, let's define the vector $\bs x=(q_x,p_x)^{\rm T}$. Then, by adopting the well-known results for conditional Gaussian systems \cite{genoni2016conditional}, we will have the resulting moments: 
\begin{align}
\begin{cases}
\bs \xi_{\rm V-B} &= \frac{1}{2(N_{\rm B}'+1)} \left(\begin{matrix}
\bs  x^{\rm T}S_{0}^{\rm T} \,  ,&\cdots &,\ \ \ \bs  x^{\rm T}S_{m-1}^{\rm T} \, 
\end{matrix}\right)^{\rm T}\\
\bs V_{\rm V-B}&=\left(\begin{matrix}
(2N_{\rm S}+1)\mathbb I_2 &0& \cdots &0\\
0& (2N_{\rm S}+1)\mathbb I_2& \cdots& \vdots    \\
\vdots& \vdots & \ddots& \vdots    \\
0& \cdots  & 0 &(2N_{\rm S}+1)\mathbb I_2 \\
\end{matrix}\right) -\frac{1}{2(N_{\rm B}'+1)}\left(\begin{matrix}
S_{0} S_{0}^{\rm T},& \cdots &,\ \ S_{0} S_{m-1}^{\rm T}\\
\vdots  & \ddots   &\vdots \\
S_{m-1} S_{0}^{\rm T},& \cdots  &,\ \ S_{m-1} S_{m-1}^{\rm T} \\
\end{matrix}\right)\\
p(\bs x )&= \frac 1 {4 (N_{\rm B}'+1) \pi }\exp\left(-\frac{|\bs x |^2}{4(N_{\rm B}'+1)}\right)
\end{cases}.\label{eq8}
\end{align} 
\end{enumerate}

\section{Performance of PA receivers in phase sensing}\label{APP:OPA}

\subsection{Quantum Crem\'er-Rao bound}

Consider a quantum state $\rho_{\bs \theta}$, which encodes a vector of parameters of interest $\bs \theta=(\theta_0,\cdots,\theta_{m-1})^{\rm T}$. Given the quantum Crem\'er-Rao theorem~\cite{helstrom1976quantum,holevo2011probabilistic,braunstein1994statistical,liu2020quantum}
%\cite{gao2014bounds,nichols2018multiparameter,vsafranek2018estimation,liu2020quantum}
, we have the following bound for multiparameter estimation: 
\begin{align}\label{CRbound}
E \ge F^{-1}
\end{align}
where $E$ is the mean-square-error (MSE) matrix with elements $\{E_{ij}=\sum_a \Tr[\rho_{\bs \theta}M_a](\hat{\theta}_i(a)-{ \theta}_i)(\hat{\theta}_j(a)-{ \theta}_j)\}$, $\{M_j\}$ refer to the positive operator-valued measure (POVM) of the measurement that satisfy the condition $\sum_a M_a=\mathbb I$ and $M_a\ge 0,\forall a$, $\hat \theta_j(a)$ is a mapping from the measurement result $a$ to the parameter $\theta_j$, $F$ is the quantum Fisher information matrix (QFIM) with the element $F_{ij}= \Tr[(L_iL_j+L_jL_i)\rho_{\bs \theta} ]/2$ and $L_i$ being the symmetric logarithmic derivative defined by $\partial \rho_{\bs \theta}/\partial \theta_i=(\rho_{\bs \theta} L_i+L_i\rho_{\bs \theta})/2$. Note that when the QFIM is a diagonal matrix,  the Crem\'er-Rao bound is achievable  \cite{helstrom1969quantum,holevo2011probabilistic}. 

Furthermore, the bound in Eq. (\ref{CRbound}) has a equivalent expression: 
\begin{align}
\Tr[W_eE]\ge  \Tr[W_e  F^{-1}]
\end{align}
where $W_e$ could be an arbitrary positive semidefinite matrix. If we choose $W_e=\mathbb I/m$ and define the average of the variances in estimating each phase as: 
\begin{align}
\epsilon=\min_{\{M_a\}} \sqrt{\frac{1}{m}\sum_{j=0}^{m-1} \sum_a \Tr[\rho_{\bs \theta}M_a](\hat{\theta}_j(a)-{ \theta}_j)(\hat{\theta}_j(a)-{ \theta}_j)},
\end{align}
we will have the Cram\'er-Rao bound for rWMSE: 
\begin{align}\label{crbound}
\epsilon\ge \sqrt{\frac{\Tr[ F^{-1}]}{m }}. 
\end{align}

\subsection{PA receivers}

Here we investigate the precision achievable by the parallel PCR scheme introduced in Section \ref{APP: OPA}. Given Eq. (\ref{eqb17}), we know that the measurement probability $\{p\left(\bs N_{{\rm D}}^{\rm tot}\right)\}$ of pPCR follows a nultivariate normal distribution. By applying the results in Ref. \cite{porat1986computation,pinele2020fisher}, we will have the element of Fisher information matrix regarding the phases:  
\begin{align}
\map F_{\theta_j,\theta_k}^{\rm ppcr}=&\sum_{\bs N_{{\rm D}}^{\rm tot}} \frac{1}{p\left(\bs N_{{\rm D}}^{\rm tot}\right)}\frac{\partial\, p\left(\bs N_{{\rm D}}^{\rm tot}\right)}{\partial \theta_j}\frac{\partial\, p\left(\bs N_{{\rm D}}^{\rm tot}\right)}{\partial \theta_k}\\
=&\frac{1 }{2}\Tr\left[\bs \Sigma^{-1}\frac{\partial \bs\Sigma}{\partial \theta_j}\bs \Sigma^{-1}\frac{\partial \bs\Sigma}{\partial \theta_k}\right]+\frac{\partial \bs \mu^T}{\partial \theta_j }\bs \Sigma^{-1}\frac{\partial \bs \mu}{\partial \theta_k },\label{eqc6}
\end{align}
where $\bs \mu:=\sqrt{2}\nu\,\bs b $ and $\bs\Sigma:=\nu\cdot\left(a\,\mathbb I_m + \bs b\bs b^T\right)$ is the covariance matrix defined in Eq. (\ref{eqb17}) with: 
\begin{align}
\bs \Sigma^{-1}&=\frac{1}{\nu a}\mathbb I_m-\frac{1}{\nu a(a+\bs b^T\bs b)}\bs b\bs b^T,\\
a&=\frac {g-1} m (N_{\rm B}'+1)(2N_{\rm S}+1) +N_{\rm S}, \\
\bs b_j&= \sqrt{\frac {2(g-1)} m  N_{\rm S}(N_{\rm S}+1)\omega_j\eta_j}\cos\theta_j.
\end{align}

Given that the first term of Eq. (\ref{eqc6}) exhibit a proportional increase with respect to $\nu$, we have: 
\begin{align}
\map F^{\rm ppcr}_{\theta_j,\theta_k}&\sim\frac{2 \nu}a \cdot \bs b_j\bs b_k\left[\delta_{jk}-\frac{\bs b_j\bs b_k}{a+\bs b^{T}\bs b}\right]\tan \theta_j\tan \theta_k. %+\mathcal O\left[ \left(\frac{(g-1)N_{\rm S}}{m^2 N_{\rm B}}\right)^2\right]
\end{align}

Next, the rWMSE may be calculated by creating the Fisher information matrix $\map F$ using the diagonal elements $\{\map F_{\theta_j}\}$ and inputting it into Equation (\ref{crbound}).  

The calculation technique for serial PCR is identical to the one described above, with the exception that the moments are $\bs \mu:=\sqrt{2}\nu\,\bs b' $ and $\bs\Sigma:=\nu\cdot\left[\textbf{diag}\left(\bs a'\right) + \bs b'\bs b^{'T}\right]$: 
\begin{align}
\bs a'_j&=g^j(g-1) (N_{\rm B}'+1)(2N_{\rm S}+1) +N_{\rm S}, \\
\bs b_j'&= \sqrt{2g^j(g-1)  N_{\rm S}(N_{\rm S}+1)\omega_j\eta_j}\cos\theta_j.
\end{align}
The corresponding Fisher information is: 
\begin{align}
\map F^{\rm spcr}_{\theta_j,\theta_k}&\sim 2 \nu \cdot  \bs b^{'T}_j \bs b^{'T}_k\left[\bs a_j^{'-1} \delta_{jk} -\frac{\bs a_j^{'-1}\bs b_j'\bs b_k'\bs a_k^{-1}}{1+\Tr\left[\bs a^{'-1}\bs b'\bs b^{'T}\right]}\right]\tan \theta_j\tan \theta_k. 
\end{align}
Therefore, one can obtain the Eq. (\ref{eq15}) in the main text. Furthermore, if we rewrite the Fisher information of Eq. (\ref{eq15}) of the main text as $\map F^{\rm pa}:=\bs A+\bs \zeta^\dag \bs \zeta$ with $\bs A$ being its first diagonal term and $\bs \zeta$ describing its second term, its inverse will be: 
\begin{align}
\Tr\left[\map F^{\rm pa -1 }\right]&= \sum_{j}\left(\bs A^{-1}\right)_{jj}+\frac{\sum_j \left(\bs A_{jj}^{-1}\bs\zeta_j\right)^2}{1+\bs b^T\bs A^{-1}\bs b-\bs\zeta^T\bs A^{-1}\bs \zeta} .
\end{align}
By applying Eq. (\ref{crbound}), one can achieve Eq. (\ref{eq9}) in Theorem \ref{theo2} of the main text.

\section{Optimal design of the CtoD protocol for phase sensing}\label{APP:advantage_phase_sensing}

In the follow-up section, we shall evaluate the achievable QFIM for the QI network with the condition $N_{\rm S}\ll N_{\rm B}$. Specifically, we first prove that the QFIM of the resulting state after implementing the heterodyne measurement in the CtoD protocol establishes a diagonal matrix. Then, we take the average of the QFIM based on the heterodyne measurement result, as the output state is projected to an orthogonal basis \cite{liu2020quantum}. Finally, we show that the classical FIM obtained by performing homodyne measurement in the last step of the CtoD can achieve the aforementioned average QFIM.

\subsection{Explicit expression of the quantum Fisher information matrix}\label{APP:QFI}

Let's compute the quantum Fisher information matrix of the conditional state $\rho_{m, \bs x}$ described by Eq. (\ref{eq8}) in terms of the phases $\{\theta_{j}|j=0,\cdots,m-1\}$. Without losing the generality, let’s have the presumption $N_{\rm B}\gg N_{\rm S}$. We will later investigate the parameter region for $N_{\rm B}=\mathcal O(N_{\rm S})$ in the practical design of measurements. On this account, the quantum Fisher information matrix (QFIM)  \cite{gao2014bounds,nichols2018multiparameter,vsafranek2018estimation,liu2020quantum} regarding different parameters is: 
\begin{align}\label{eqd1}
\map F_{ij}=\frac 1 2 \map R^{-1}_{\alpha\beta,\mu\nu}\frac{\partial  \bs V_{\alpha\beta}}{\partial \theta_{ i}}\frac{\partial   \bs V_{\mu\nu}}{\partial \theta_j} + \bs V_{\mu\nu}^{-1} \frac{\partial \bs \xi_{\mu}}{\partial \theta_{ i}}\frac{\partial \bs \xi_{\nu}}{\partial \theta_j},
\end{align}
where $\map R=\bs V\otimes \bs V + \bs \Omega\otimes \bs \Omega/4$. 

Further, we can define the matrix $\map R_0=\bs V_0\otimes \bs V_0 + \bs \Omega\otimes \bs \Omega/4$ with $\bs V_0$ being the first term of Eq. (\ref{eq8}). Then, given the relation $\map R-\map R_0\propto \mathcal O(N_{\rm S}^3/N_{\rm B})$, we have the condition $\lim_{n\to \infty }(\mathbb I-\map R^{-1}_0 \map R)^n=0$ \cite{horn2012matrix} as well as the following relation: 
\begin{align}
\map R^{-1}=&\sum_{n=0}^{\infty} \left[\map R_0^{-1}\left(\map R_0-\map R\right)\right]^n \map R_0^{-1}\\
=&\map R_0^{-1} +\map E(N_{\rm S},N_{\rm B}),
\end{align}
where $\map E(N_{\rm S},N_{\rm B})$ is an error matrix that has the scaling $ \mathcal O\left[\left(m_{\rm re}N_{\rm B}N_{\rm S}\right)^{-1}\right]$. Therefore, given the following relations: 
\begin{align}
\map R_0^{-1}&= \mathbb I_m\otimes \mathbb I_m\otimes \left(\begin{matrix}
r_a&0&0&-r_{\rm B}\\
0&r_a&r_{\rm B}&0\\
0&r_{\rm B}&r_a&0\\
-r_{\rm B}&0&0&r_a
\end{matrix}\right),\\
\frac{\partial \bs V_{\rm V}}{\partial \theta_j}&=-\frac{1}{2(N_{\rm B}'+1)}\left(\begin{matrix}
0,& \cdots &S_{0}\frac{\partial S_{j}^{\rm T}}{\partial \theta_j} &\cdots&,\ \ 0\\
\vdots  &0& \vdots   &0&\vdots \\
\frac{\partial S_{m-1}}{\partial \theta_j} S_{0}^{\rm T},& \cdots  &,\ \ S_{j}\frac{\partial S_{j}^{\rm T}}{\partial \theta_j} +\frac{\partial S_{j}}{\partial \theta_j} S_{j}^{\rm T} &\cdots&S_{j}\frac{\partial S_{m-1}^{\rm T}}{\partial \theta_j}\\
\vdots,&0& \vdots  &0&,\ \ \vdots \\
0,& \cdots & S_{m-1}\frac{\partial S_{j}^{\rm T}}{\partial \theta_j}&\cdots&,\ \ 0 
\end{matrix}\right),\\
\bs V_{\rm V}^{-1}&\nonumber=\left(\begin{matrix}
(2N_{\rm S}+1)^{-1}\mathbb I_2 &0& \cdots &0\\
0& (2N_{\rm S}+1)^{-1}\mathbb I_2& \cdots& \vdots    \\
\vdots& \vdots & \ddots& \vdots    \\
0& \cdots  & 0 &(2N_{\rm S}+1)^{-1}\mathbb I_2 \\
\end{matrix}\right)\\ &+\frac{1}{2(2N_{\rm S}+1)^2(N_{\rm B}'+1)-4(2N_{\rm S}+1)N_{\rm S}(N_{\rm S}+1)C_\omega}\left(\begin{matrix}
S_{0} S_{0}^{\rm T},& \cdots &,\ \ S_{0} S_{m-1}^{\rm T}\\
\vdots  & \ddots   &\vdots \\
S_{m-1} S_{0}^{\rm T},& \cdots  &,\ \ S_{m-1} S_{m-1}^{\rm T} \\
\end{matrix}\right),\label{eq14a}
\end{align}
where $r_a=\frac{16(2N_{\rm S} + 1)^2}{256 N_{\rm S}^4 + 512N_{\rm S}^3 + 384 N_{\rm S}^2 + 128N_{\rm S} + 15}$,  $r_{\rm B}=-\frac{4}{256N_{\rm S}^4 + 512N_{\rm S}^3 + 384N_{\rm S}^2 + 128N_{\rm S} + 15}$, $C_\omega=\sum_j \omega_j\eta_j$, the overall Fisher information matrix is lower bounded as follows: 
\begin{align}
F= & \frac{N_{\rm S}(N_{\rm S}+1)}{(N_{\rm B}'+1)^2(2N_{\rm S}+1)}|\bs x |^2\textbf{ diag}\{\omega_0\eta_0\cdots,\omega_{m-1}\eta_{m-1} \} \nonumber \\&+ \frac{2N_{\rm S}^2(N_{\rm S}+1)^2}{[(2N_{\rm S}+1)^2(N_{\rm B}'+1)-2(2N_{\rm S}+1)N_{\rm S}(N_{\rm S}+1)C_\omega](N_{\rm B}'+1)^2}  F'+  \map E'(N_{\rm S},N_{\rm B}),
\end{align}
where $F'$ is a matrix with the elements $F'_{ij}=\omega_{i}\omega_j\eta_{i}\eta_j \bs x^{\rm T} \dot{\mathbb R}_{i}^{\rm T} \mathbb R_{i} \mathbb R_j^{\rm T} \dot{\mathbb R}_j \bs x\equiv \omega_{i}\omega_j\eta_{ i}\eta_j| \bs x|^2$, $\map E'(N_{\rm S},N_{\rm B})$ has the scaling $\mathcal O\left({N_{\rm S}^3}/({N_{\rm B}m_{\rm re})^3}\right)$.  

Therefore, the average Fisher information matrix is: 
\begin{align}
\overline F=&\int \frac{\d^2 \bs x}{\pi}p(\bs x )F\\
\label{eq17}=&\frac{4N_{\rm S}(N_{\rm S}+1)}{(N_{\rm B}'+1)(2N_{\rm S}+1)} \textbf{ diag}\{\omega_0\eta_0\cdots,\omega_{m-1}\eta_{m-1} \}\nonumber \\
& + \frac{8N_{\rm S}^2(N_{\rm S}+1)^2}{[(2N_{\rm S}+1)^2(N_{\rm B}'+1)-2(2N_{\rm S}+1)N_{\rm S}(N_{\rm S}+1)\bs e^{\rm T}\bs v ](N_{\rm B}'+1)}  \bs v \bs v^{\rm T}+  \map E'(N_{\rm S},N_{\rm B}),
\end{align}
where $\bs v$ is a vector $\bs v= (\omega_0\eta_0,\cdots,\omega_{m-1}\eta_{m-1} )$ and $\bs e=(1,\cdots,1)^{\rm T}$, the first and second term of Eq. (\ref{eq17}) has the scaling $\mathcal O(N_{\rm S}/(N_{\rm B}m_{\rm re}))$ and $\mathcal O(N_{\rm S}^2/(N_{\rm B}m_{\rm re})^2)$, respectively. The dominating part of the average Fisher information matrix $\overline F$ is diagonal, which indicates that it is achievable by independent single-parameter estimation protocols \cite{helstrom1969quantum,holevo2011probabilistic}. In addition, the first term of Eq. (\ref{eq17}) has the same scaling as that shown in the single-parameter estimation \cite{gagatsos2017bounding}. 

\subsection{Repetition of experiments}\label{APP:Qrepetition}

The average Fisher information matrix in Eq. (\ref{eq17}) is derived by taking an average of the Fisher information matrix for conditional states, while in practice, different rounds of experiments will lead to different conditional states. Consider a scenario where the experimenter conducts the CtoD protocol with $\nu$ rounds of experiment. The resulting state will be: 
\begin{align}
\rho_{{\rm V},\nu}=\bigotimes_{n=0}^{\nu-1} \rho_{{\rm V},\bs x_n} 
\end{align}  
which has the moments: 
\begin{align}
\begin{cases}
\bs \xi_{{\rm V},\nu} &= \frac{1 }{2(N_{\rm B}'+1)} \left(\begin{matrix}
\bs  x_0^{\rm T}S_{0}^{\rm T} \,  ,&\cdots &,\ \ \  \bs x_0^{\rm T}S_{m-1}^{\rm T} \,,\cdots,\, \bs  x_{m-1}^{\rm T}S_{0}^{\rm T}  ,&\cdots &,\ \ \ \bs x_{m-1}^{\rm T} S_{m-1}^{\rm T} \, 
\end{matrix}\right)^{\rm T} \\
\bs V_{{\rm V},\nu}&=\left(\begin{matrix}
(2N_{\rm S}+1)\mathbb I_2 &0& \cdots &0\\
0&  (2N_{\rm S}+1)\mathbb I_2& \cdots& \vdots    \\
\vdots& \vdots & \ddots& \vdots    \\
0& \cdots  & 0 & (2N_{\rm S}+1)\mathbb I_2 \\
\end{matrix}\right)\otimes \mathbb I_\nu -\frac{ 1}{2(N_{\rm B}'+1)}\left(\begin{matrix}
S_{0} S_{0}^{\rm T},& \cdots &,\ \ S_{0} S_{m-1}^{\rm T}\\
\vdots  & \ddots   &\vdots \\
S_{m-1} S_{0}^{\rm T},& \cdots  &,\ \ S_{m-1} S_{m-1}^{\rm T} \\
\end{matrix}\right)\otimes \mathbb I_\nu\\
p(\{\bs x_n \})&= \frac 1 {(4 (N_{\rm B}'+1))^\nu }\exp\left(-\frac{\sum_n|\bs x_n |^2}{4(N_{\rm B}'+1)}\right)
\end{cases}
\end{align}
where $\bs x_n$ refer to the heterodyne measurement reuslt from $n$-th repetition of the steps (I)-(IV) and (V.B). 

\begin{enumerate}
\item [(VI.B)]\emph{Phase rotation and displacement concentration:---} The experimenter can implement single-mode phase-rotation operations based on the measurement results $\{\bs x_n\}$ to make the phase of displacement equal to that of the first mode. Then, the experimenter can implement $m$ $\nu$-mode beam-splitter operations in the redundant space to concentrate the displacement to $m$ modes with moments: 
\begin{align}\label{eq20}
\begin{cases}
\bs \xi_{{\rm VI},\nu} &= \frac{\sqrt{\sum_j |\bs x_j|^2} }{2(N_{\rm B}'+1)} \left(\begin{matrix}
\bs  e_x^{\rm T}S_{0} ^{\rm T}\,  ,&\cdots &,\ \ \ \bs e_x^{\rm T}S_{m-1} ^{\rm T}\, 
\end{matrix}\right)^{\rm T} \\
\bs V_{{\rm VI},\nu}&=\left(\begin{matrix}
(2N_{\rm S}+1)\mathbb I_2 &0& \cdots &0\\
0&  (2N_{\rm S}+1)\mathbb I_2& \cdots& \vdots    \\
\vdots& \vdots & \ddots& \vdots    \\
0& \cdots  & 0 & (2N_{\rm S}+1)\mathbb I_2 \\
\end{matrix}\right)-\frac{ 1}{2(N_{\rm B}'+1)}\left(\begin{matrix}
S_{0} S_{0}^{\rm T},& \cdots &,\ \ S_{0} S_{m-1}^{\rm T}\\
\vdots  & \ddots   &\vdots \\
S_{m-1} S_{0}^{\rm T},& \cdots  &,\ \ S_{m-1} S_{m-1}^{\rm T} \\
\end{matrix}\right)\\
p(\{\bs x_k \})&= \frac 1 {(4 (N_{\rm B}'+1))^\nu }\exp\left(-\frac{\sum_k|\bs x_k |^2}{4(N_{\rm B}'+1)}\right)
\end{cases}
\end{align}
where $\bs e_x= \bs x_0/ |\bs x_0|$.
\end{enumerate}

Given the relations $\bs \xi_{{\rm VI},\nu} =( \sqrt{\sum_j |\bs x_j|^2}/|\bs x_0|)\bs \xi_{{\rm V},\nu} $ and $\bs V_{{\rm VI},\nu}=\bs V_{{\rm V},\nu}$, we can quickly derive the quantum Fisher information matrix: 
\begin{align}
\overline F_\nu=&\frac{4\nu N_{\rm S}(N_{\rm S}+1)}{(N_{\rm B}'+1)(2N_{\rm S}+1)} \textbf{ diag}\{\omega_0\eta_0\cdots,\omega_{m-1}\eta_{m-1} \}\nonumber \\
& + \frac{8\nu N_{\rm S}^2(N_{\rm S}+1)^2}{[(2N_{\rm S}+1)^2(N_{\rm B}'+1)-2(2N_{\rm S}+1)N_{\rm S}(N_{\rm S}+1)\bs e^{\rm T}\bs v ](N_{\rm B}'+1)}  \bs v \bs v^{\rm T}+  \map E'(N_{\rm S},N_{\rm B}),
\end{align}
where the expressions for $\bs \nu$ and $\bs e$ are given in Eq. (\ref{eq17}).

\subsection{Optimal design of the estimation protocol $(N_{\rm B}>N_{\rm S})$}\label{APP:OEP}

Let's consider homodyne measurement on the conditional state in step V.B, which produces the probability distribution  \cite{serafini2017quantum}: 
\begin{align}\label{eqc15}
p(\bs q )
&=\frac{\exp\left[-\frac 1 2 (\bs q -\bs \xi_{{\rm V},q})^{\rm T} \left[(2N_{\rm S}+1)\mathbb I_m-\frac{2N_{\rm S}(N_{\rm S}+1)}{N_{\rm B}'+1}T_{\bs \theta}\right]^{-1}(\bs q -\bs \xi_{{\rm V},q})  \right]}{\sqrt{\prod_{j=0}^{m-1} 2\pi \left(2N_{\rm S}+1-\frac{2N_{\rm S}(N_{\rm S}+1)\omega_j\eta_j}{N_{\rm B}'+1}\right)}},
\end{align}
where $T_{\bs\theta}$ is a $m\times m $ matrix with matrix elements $T_{{\bs\theta,ij}}=\sqrt{\omega_{ i} \omega_j \eta_{ i} \eta_j} \cos (\theta_{ i}-\theta_j)$, $\bs q:=(\hat q_0,\cdots,\hat q_{m-1})^{\rm T}$ is a real vector with $m$ elements regarding to the measurement result, $\bs \xi_{{\rm V},q}=\sqrt{N_{\rm S}(N_{\rm S}+1)}/(N_{\rm B}'+1)[\sqrt{\omega_{0}\eta_0}(q_x\cos \theta_0+p_x\sin\theta_0),\cdots, \sqrt{\omega_{m-1}\eta_{m-1}}(q_x\cos \theta_{m-1}+p_x\sin\theta_{m-1})]^{\rm T}$ is the position component of $\bs \xi_{{\rm V}}$. 

Therefore, we have the classical Fisher information matrix: 
\begin{align}
&\nonumber F_{mn}\\
=& \lim_{\epsilon_m\to 0\atop \epsilon_n\to 0} \frac{8\left[1-\int \d \bs q \sqrt{ p_{\bs \theta}(\bs q )p_{\bs \theta+\epsilon_m+\epsilon_n}(\bs q)}\right]}{\epsilon_m\epsilon_n} \\
\ge & \lim_{\epsilon_m\to 0\atop \epsilon_n\to 0} \frac{8\left[1-\frac{\exp\left[-\frac 1 {16} \left(\bs\xi_{{\rm V},q}-\bs\xi_{{\rm V},q,\epsilon_m,\epsilon_n}\right)^{\rm T}\left(M_{\bs \theta}+M_{\bs \theta+\epsilon_m,\epsilon_n}\right)\left(\bs\xi_{{\rm V},q}-\bs\xi_{{\rm V},q,\epsilon_m,\epsilon_n}\right)\right]}{\sqrt{\det\left\{\frac 1 2 \mathbb I_m+\frac 1 2 M_{\bs \theta}^{-1}M_{\bs \theta+\epsilon_m,\epsilon_n}\right\}}}\right]}{\epsilon_m\epsilon_n}\label{eqd16}\\
=& \lim_{\epsilon_m\to 0\atop \epsilon_n\to 0} \frac{8\left\{1-\exp\left[-\frac 1 {8} \left(\bs\xi_{{\rm V},q}-\bs\xi_{{\rm V},q,\epsilon_m,\epsilon_n}\right)^{\rm T}M_{\bs \theta}\left(\bs\xi_{{\rm V},q}-\bs\xi_{{\rm V},q,\epsilon_m,\epsilon_n}\right)\right]\right\}}{\epsilon_m\epsilon_n}\\
=&\,\,\,2\frac{\partial{\bs\xi}_{{\rm V},q}^{\rm T}}{\partial \theta_m}M_{\bs \theta}\frac{\partial{\bs\xi}_{{\rm V},q}}{\partial \theta_n},\label{eqb23}
\end{align}
where $\xi_{\epsilon_m,\epsilon_n}$ and $\bs \theta+\epsilon_m+\epsilon_n$ refer to case where the $m$-th and $n$-th phases have negligible additional parts $\epsilon_m$ and $\epsilon_n$, respectively, $M_{\bs \theta}=\left[(2N_{\rm S}+1)\mathbb I_m-2N_{\rm S}(N_{\rm S}+1)/(N_{\rm B}'+1)T_{\bs \theta}\right]^{-1}$, and Eq. (\ref{eqd16}) is obtained by the relation: 
\begin{align}
\nonumber &\int \d \bs q \exp\left\{-\frac 1 4 (\bs q -\bs \xi_{{\rm V},q})^{\rm T} M_{\bs \theta}(\bs q -\bs \xi_{{\rm V},q})-\frac 1 4 (\bs q -\bs \xi_{{\rm V},q,\epsilon_m,\epsilon_n})^{\rm T} M_{\bs \theta+\epsilon_m,\epsilon_n}(\bs q -\bs \xi_{{\rm V},q,\epsilon_m,\epsilon_n})\right\}\\
=&\int \d \bs q \exp\left\{-\frac 1 4 \bs q ^{\rm T} M_{\bs \theta}\bs q -\frac 1 4 (\bs q -\bs \xi_{{\rm V},q,\epsilon_m,\epsilon_n}+\bs \xi_{{\rm V},q})^{\rm T} M_{\bs \theta+\epsilon_m,\epsilon_n}(\bs q -\bs \xi_{{\rm V},q,\epsilon_m,\epsilon_n}+\bs \xi_{{\rm V},q})\right\}\\
=&\int \d \bs q \exp\left\{ -\frac 1 4 (-\bs q -\bs \xi_{{\rm V},q}+\bs \xi_{{\rm V},q,\epsilon_m,\epsilon_n})^{\rm T} M_{\bs \theta}(-\bs q -\bs \xi_{{\rm V},q}+\bs \xi_{{\rm V},q,\epsilon_m,\epsilon_n})-\frac 1 4 \bs q ^{\rm T} M_{\bs \theta+\epsilon_m,\epsilon_n}\bs q\right\}\\
\ge \nonumber &\int \d \bs q \exp\left\{ -\frac 1 4 \bs q ^{\rm T} \left(M_{\bs \theta}+M_{\bs \theta+\epsilon_m,\epsilon_n}\right)\bs q  -\frac 1 8 \bs q ^{\rm T} \left(M_{\bs \theta}+M_{\bs \theta+\epsilon_m,\epsilon_n}\right)( \bs \xi_{{\rm V},q}-\bs \xi_{{\rm V},q,\epsilon_m,\epsilon_n})\right.\\
&\left. -\frac 1 8 (\bs \xi_{{\rm V},q}-\bs \xi_{{\rm V},q,\epsilon_m,\epsilon_n})^{\rm T} \left(M_{\bs \theta}+M_{\bs \theta+\epsilon_m,\epsilon_n}\right)\bs q -\frac 1 8(\bs \xi_{{\rm V},q}-\bs \xi_{{\rm V},q,\epsilon_m,\epsilon_n})^{\rm T} \left(M_{\bs \theta}+M_{\bs \theta+\epsilon_m,\epsilon_n}\right)( \bs \xi_{{\rm V},q}-\bs \xi_{{\rm V},q,\epsilon_m,\epsilon_n}) \right\}\\
= &\sqrt{\frac{\pi^m}{\det \left(\frac1 4M_{\bs \theta}+\frac1 4M_{\bs \theta+\epsilon_m,\epsilon_n}\right)}}  \exp\left\{ -\frac 1 {16} ( \bs \xi_{{\rm V},q,\epsilon_m,\epsilon_n}-\bs \xi_{{\rm V},q})^{\rm T} \left(M_{\bs \theta}+M_{\bs \theta+\epsilon_m,\epsilon_n}\right)( \bs \xi_{{\rm V},q,\epsilon_m,\epsilon_n}-\bs \xi_{{\rm V},q})\right\}.
\end{align}
Here the inequality is derived from the Cauchy–Schwarz inequality: $\sum_n x_ny_n\le \sum_n x_n^2 \sum_m y_m^2$.

Here we have the relation: 
\begin{align}
&\left[(2N_{\rm S}+1)\mathbb I_m-\frac{2N_{\rm S}(N_{\rm S}+1)}{N_{\rm B}'+1}T_{\bs \theta}\right]^{-1}=\frac{1}{2N_{\rm S}+1}\mathbb I_m +\frac{(2N_{\rm S}+1)(\bs s\bs s^{\rm T}+\bs c\bs c^{\rm T})+\bs s\bs s^{\rm T}\bs c\bs c^{\rm T}+\bs c\bs c^{\rm T}\bs s\bs s^{\rm T}-\bs c^{\rm T} \bs c\bs s\bs s^{\rm T}-\bs s^{\rm T} \bs s \bs c\bs c^{\rm T}}{(2N_{\rm S}+1)\left[(2N_{\rm S}+1-\bs s^{\rm T} \bs s)(2N_{\rm S}+1-\bs c^{\rm T} \bs c) -|\bs c^{\rm T} \bs s |^2\right]},
\end{align}
where $\bs c=\sqrt{2N_{\rm S}(N_{\rm S}+1)/(N_{\rm B}'+1)}(\sqrt{\omega_0\eta_0}\cos(\theta_0),\cdots,\sqrt{\omega_{m-1}\eta_{m-1}}\cos(\theta_{m-1}))$ and $\bs s=\sqrt{2N_{\rm S}(N_{\rm S}+1)/(N_{\rm B}'+1)}(\sqrt{\omega_0\eta_0}\sin(\theta_0),\cdots,\sqrt{\omega_{m-1}\eta_{m-1}}\sin(\theta_{m-1}))$, the second term has the scaling $\mathcal O\left(1/(mN_{\rm B})\right)$ for $1<N_{\rm S}\ll N_{\rm B}$ and $\mathcal O\left(N_{\rm S}/(mN_{\rm B})\right)$ for $N_{\rm S}\ll1\ll N_{\rm B}$. Here, we assume the condition $\omega_j\sim \mathcal O(1/m),\forall j$. Then, one can compute the explicit expression for the Fisher information: 
\begin{align}
F
&= \frac{2N_{\rm S}(N_{\rm S}+1)}{(N_{\rm B}'+1)^2(2N_{\rm S}+1)}
\textbf{ diag}\{\omega_0\eta_0(-q_x\cdots),\cdots,\omega_{m-1}\eta_{m-1}(-q_x\sin\theta_{m-1}+p_x\cos\theta_{m-1})^2 \} +\map F'(N_{\rm S},N_{\rm B}),
\end{align}
where $\map F'(N_{\rm S},N_{\rm B})$ has the scaling $\mathcal O\left({N_{\rm S}^2(q_x+p_x)^2}/(m^2_{\rm re}{N_{\rm B}^3)}\right)$ for either the case $1<N_{\rm S}\ll N_{\rm B}$ or the case  $N_{\rm S}\ll1\ll N_{\rm B}$.

If we adopt the protocol VI in $\nu$ rounds of experiment and integrate over the set of measurement results $\{\bs x_n\}$, we have: 
\begin{align}
{\overline{F}}&=\int \prod_j\d \bs x_j p(\{\bs x_j\}) \nonumber \\
&\times \frac{2N_{\rm S}(N_{\rm S}+1)}{(N_{\rm B}'+1)^2(2N_{\rm S}+1)}
\textbf{ diag}\left\{\omega_0\eta_0\left(\sum_{l=0}^{\nu-1}q^2_{x,l}\right.\cdots,\omega_{m-1}\eta_{m-1}\left(\sum_k q_{x,k}^2\sin^2\theta_{m-1}+p_{x,k}^2\cos^2\theta_{m-1}\right) \right\} +\map F'(N_{\rm S},N_{\rm B})\\
&=\label{eqb26} \frac{4\nu N_{\rm S}(N_{\rm S}+1)}{(N_{\rm B}'+1)(2N_{\rm S}+1)}\textbf{ diag}\{\omega_0\eta_0\cdots,\omega_{m-1}\eta_{m-1} \}+  \map F''(N_{\rm S},N_{\rm B}), 
\end{align}
where $ \map F''(N_{\rm S},N_{\rm B})$ has the scaling $\mathcal O\left(\nu{N_{\rm S}^2}/(m^2_{\rm re}{N_{\rm B}^2})\right)$. Therefore, for either the case $1<N_{\rm S}\ll N_{\rm B}$ or the case  $N_{\rm S}\ll1\ll N_{\rm B}$, we have the average of the estimation variance for $\{\theta_j\}$: 
\begin{align}
\overline E&=\frac 1 {m} \sum_{j=0}^{m-1} \left\<(\hat \theta-\overline \theta)^2\right\>\\
&=\frac 1 {m} \sum_{j=0}^{m-1} \overline F_{jj}^{-1} + \mathcal O\left(\frac{N_{\rm S}}{\nu N_{\rm B}}\right)\\
&= \frac{(N_{\rm B}'+1)(2N_{\rm S}+1)}{4m\nu N_{\rm S}(N_{\rm S}+1)}\sum_{j=0}^{m-1}(\omega_j\eta_j)^{-1}+\mathcal O\left(\frac{1}{\nu }\right),
\end{align}
where the second equation is obtained by taking into consideration the effect of nuisance parameters \cite{suzuki2020quantum}. Finally, by defining the standard derivation by $\epsilon=\sqrt{\overline E}$, we can achieve the Eq. (\ref{wrmse}) in the main text. 

Note that the achievable rWMSE can also be lower bounded by that obtained by a favorable case where only one of the $m$ phases is unknown \cite{suzuki2020quantum}. In this case, the scaling of the lower bound regarding $N_{\rm S}$, $N_{\rm B}$, and $\nu$ can be given by an RMSE for estimating identical phases. It has been shown in Ref. \cite{shi2022fulfilling} that the optimal RMSE achieves the scaling $\map O(\sqrt{N_{\rm B}/(\nu N_{\rm S})})$. Therefore, we can conclude that Eq. (\ref{wrmse}) attains the optimal scaling regarding $N_{\rm S}$, $N_{\rm B}$, and $\nu$, despite noncommutative generators for alternative phases.

\section{CtoD protocol for Pattern Classification}\label{APP:PC}

\subsection{Set-up}

Let's consider the scenario where the QI network is employed to discriminate the following two hypotheses: 
\begin{align}
\begin{cases}
\text{Hypotheses I:\ \ \ } \bs \eta=\bs \eta^{(0)}\\
\text{Hypotheses II:\ \ } \bs \eta=\bs \eta^{(1)}
\end{cases}
\end{align}
where $\bs \eta^{(h)}:=( \eta^{(h)}_0,\cdots, \eta^{(h)}_{m-1})^{\rm T}$ for $h=0,1$ refers to the vector of reflectivity encoded from the output state $\rho_{\bs \eta^{(h)}}$ of the QI network (see Eq. (\ref{CoV}) in the main text). Following the first and second steps of the CtoD protocol with $\nu$ rounds of experiment in Appendix \ref{APP:Qrepetition}, it is possible to produce an $m$-mode conditional state described by Eq. (\ref{C2D-Dis-Cov}) in the main text: 
\begin{align}
\rho_{\bs \eta^{(h)},{\rm pe}}
&=  \rho_{\bs \eta^{(h)},\sqrt \nu\overline{\bs x}}\otimes \rho_{\bs \eta^{(h)},{\bs 0}}^{\otimes \nu-1}, \ \ \ h=0,1
\end{align}
except for its displacement where  $\overline{\bs x}= \sqrt{\sum_{n=0}^{\nu-1}|\bs x_n|^2/\nu}$, where the subscript `pe' stands for parameter estimation. Here, we introduce an alternative second step of CtoD for hypothesis testing:  
\begin{enumerate}
\item [(2.HP)] Use the $m$ $\nu$-mode beam-splitter operations to uniformly distributed the displacements to the $\nu$ $m$-mode states:   
\begin{align}
\rho_{\bs \eta^{(h)},{\rm hp}}
&=  \rho_{\bs \eta^{(h)},\overline{\bs x}}^{\otimes \nu}, \ \ \ h=0,1, 
\end{align}
where the subscript `hp' stands for hypothesis testing.
\end{enumerate}

\subsection{The quantum Chernoff bound}

Through the step mentioned by the last subsection, we can apply the quantum Chernoff bound for discriminating identical states and bound the corresponding error \cite{audenaert2007discriminating,pirandola2008computable}: 
\begin{align}
p_{\rm hp} \left(\rho_{\bs \eta^{(0)}}^{\otimes \nu},\rho_{\bs \eta^{(1)}}^{\otimes \nu}\right)\le &\int \prod_n \d \bs x_n p(\{\bs x_n\}) p_{\rm hp} \left(\rho_{\bs \eta^{(0)},\overline{\bs x}}^{\otimes \nu},\rho_{\bs \eta^{(1)},\overline{\bs x}}^{\otimes \nu}\right)\\
p_{\rm hp} \left(\rho_{\bs \eta^{(0)},\overline{\bs x}}^{\otimes \nu},\rho_{\bs \eta^{(1)},\overline{\bs x}}^{\otimes \nu}\right):=&1-\frac 1 2 \max_{\{\hat \Pi_{h,\overline{\bs x}} \}} \Tr\left[\hat \Pi_{h,\overline{\bs x}}\rho_{\bs \eta^{(h)},\overline{\bs x}}^{\otimes \nu} \right]\\
\le& \frac 1 2 \left(\inf_{s\in[0,1]}\Tr\left[\rho_{\bs \eta^{(0)},\overline{\bs x}}^s\rho_{\bs \eta^{(1)},\overline{\bs x}}^{1-s} \right]\right)^\nu\label{eq37}
\end{align}
where the inequality is derived from the quantum data processing inequality after heterodyne measurement  \cite{nielsen2010quantum}, $\{\hat \Pi_{j,\overline{\bs x}}\}$ refers to an arbitrary $(m\nu)$-mode measurement. In particular, we can further use the relations for Gaussian quantum systems  \cite{pirandola2008computable}: 
\begin{align}
\Tr\left[\rho_{\bs \eta_0,\overline{\bs x}}^s\rho_{\bs \eta_1,\overline{\bs x}}^{1-s}\right]&=  Q_s \exp\left\{-\frac 1 2 (\overline{\bs \xi}_{\bs \eta^{(0)}}-\overline{\bs \xi}_{\bs \eta^{(1)}})^{\rm T}\left( {\bs V}_{\bs \eta^{(0)},s}+ {\bs V}_{\bs \eta^{(1)},1-s}\right)^{-1}(\overline{\bs \xi}_{\bs \eta^{(0)}}-\overline{\bs \xi}_{\bs \eta^{(1)}})\right\}\\
Q_s&= \frac{2^m \prod_{j=0}^{m-1} G_{s} (\mu_j^{(0)})G_{1-s} (\mu_j^{(1)})}{\sqrt{\det \left[ {\bs V}_{\bs \eta^{(0)},s}+ {\bs V}_{\bs \eta^{(1)},1-s}\right]}}\\
{\bs V}_{\bs \eta^{(0)},s}&= S_{{\rm sym}}^{(h)} \left[\bigoplus_{j=0}^{m-1} \Lambda_{s } (\mu_j^{(h)})\mathbb I_2 \right] S_{{\rm sym}}^{(h){\rm T}}
\end{align}
where $S^{(h)}_{{\rm sym}}$ is a symplectic transformation defined by ${\bs V}_{\bs \eta_h}=S^{(h)}_{{\rm sym}} \left(\bigoplus_{j=0}^{m-1} \mu_j^{(h)}\mathbb I_2 \right)S^{(h){\rm T}}_{{\rm sym}}$ with $\{\mu^{(h)}_j\}$ being the symplectic eigenvalues, with $\bs V_{\bs \eta_h}$ being the covariance matrix defined in Eq. (\ref{C2D-Dis-Cov}) of the main text, the displacement vector $\overline{\bs \xi}_{\bs \eta_h}$ is defined by $\overline{\bs \xi}_{\bs \eta_h}:= 1/(2(N_{\rm B}'+1)) \left(
S_{0} \, \overline{\bs  x} ,\cdots ,\ \ \ S_{m-1} \, \overline{\bs x}\right)^{\rm T}$, $\{{\rm S}_j\}$ are defined in Eq. (\ref{CoV}) in the main text.

Without losing the generality, let's consider the practical situation where the phases are known. Then, one can implement phase correction operations to have $\theta_j^{(h)}=0$ for 
$j=0,\cdots,m-1,h=0,1$. In this case, the covariance matrices has the form $\bs V_{\bs \eta_h}=(2N_{\rm S}+1)\mathbb I_{2m} -2N_{\rm S}(N_{\rm S}+1)/(N_{\rm B}'+1)\bs \eta_\omega^{(h)}\bs \eta_\omega^{(h)T}\otimes \mathbb I_2$ with $\bs \eta_\omega^{(h)} = \left(\sqrt{\omega_{0}\eta^{(h)}_{0}},\cdots,\sqrt{\omega_{m_1}\eta^{(h)}_{m-1}} \right)^{\rm T}$, we could have: 
\begin{align}
\begin{cases}
\mu^{(h)}_0&=2N_{\rm S}+1-\frac{2N_{\rm S}(N_{\rm S}+1)}{N_{\rm B}'+1}\bs \eta_\omega^{(h){\rm T}} \bs \eta_\omega^{(h)}  \\
&=2N_{\rm S}+1-\mathcal O\left(\frac{N_{\rm S}^2}{m_{\rm re}N_{\rm B}}\right) \\
\mu^{(h)}_j&=2N_{\rm S}+1,\ \ \ \ j=1,\cdots m-1
\end{cases}
\end{align}
Thereby, if we take into consideration the condition $N_{\rm S}=\mathcal O(1)\ll N_{\rm B}$, we have: 
\begin{align}
{\bs V}_{\bs \eta_h,s}=& \Lambda_{s}(2N_{\rm S}+1) \mathbb I_{2m}+\frac{\Lambda_{ s}(\mu_0^{(h)})-\Lambda_{ s}(2N_{\rm S}+1)}{\bs \eta_\omega^{(h){\rm T}} \bs \eta_\omega^{(h)}} \bs \eta_\omega^{(h)}\bs \eta_\omega^{(h){\rm T}} \otimes \mathbb I_2\\
=&\Lambda_{s}(2N_{\rm S}+1) \mathbb I_{2m}-\frac{8N_{\rm S}(N_{\rm S}+1)s(2N_{\rm S}+2)^{s-1}(2N_{\rm S})^{s-1}}{(N_{\rm B}'+1)[(2N_{\rm S}+2)^s-(2N_{\rm S})^s]^2}\bs \eta_\omega^{(h)}\bs \eta_\omega^{(h){\rm T}} \otimes \mathbb I_2+ \mathcal O\left(\frac{N_{\rm S}^2 \epsilon_d }{N_{\rm B}}\right)\bs \eta_\omega^{(h)}\bs \eta_\omega^{(h){\rm T}} \otimes \mathbb I_2\\
=&\Lambda_{s}(2N_{\rm S}+1) \mathbb I_{2m}+ \mathcal O\left(\frac{ 1}{N_{\rm B}}\right)\bs \eta_\omega^{(h)}\bs \eta_\omega^{(h){\rm T}}\otimes \mathbb I_2
\end{align}
where $\epsilon_d$ is a constant from the definition of the derivative limit \cite{swokowski1979calculus}. 

With these premises, we have the relation: 
\begin{align}
\Tr\left[\rho_{\bs \eta_0}^s\rho_{\bs \eta_1}^{1-s}\right]\le &  \exp\left\{-\frac 1 2 (\overline{\bs \xi}_{\bs \eta^{(0)}}-\overline{\bs \xi}_{\bs \eta^{(1)}})^{\rm T}\left( {\bs V}_{\bs \eta^{(0)},s}+ {\bs V}_{\bs \eta^{(1)},1-s}\right)^{-1}(\overline{\bs \xi}_{\bs \eta^{(0)}}-\overline{\bs \xi}_{\bs \eta^{(1)}})\right\}\\
= & \exp\left\{-\frac 1 2 \frac{|\overline{\bs \xi}_{\bs \eta^{(0)}}-\overline{\bs \xi}_{\bs \eta^{(1)}}|^2}{\Lambda_{ s}(2N_{\rm S}+1)+\Lambda_{1-s}(2N_{\rm S}+1)}+\mathcal O \left(\frac{m^2(\bs x_q+\bs x_p)^2}{m_{\rm re}^2N_{\rm B}^3}\right)\right\} \\
=&\exp\left\{-\frac {N_{\rm S}(N_{\rm S}+1)} {2(N_{\rm B}'+1)^2} \frac{\sum_{j=0}^{m-1} \omega_j\left(\sqrt{\eta_j^{(0) }}-\sqrt{\eta_j^{(1) }}\right)^2}{\Lambda_{ s}(2N_{\rm S}+1)+\Lambda_{1-s}(2N_{\rm S}+1)}|\overline{\bs x}|^2\right\} + \mathcal O\left(\frac{m^2(\bs x_q+\bs x_p)^2 }{m_{\rm re}^2N_{\rm B}^3}\right)
\end{align}

Finally, the discrimination error is  bounded by:
\begin{align}
p_{\rm hp} \left(\rho_{\bs \eta^{(0)},\overline{\bs x}}^{\otimes \nu},\rho_{\bs \eta^{(1)},\overline{\bs x}}^{\otimes \nu}\right)
\le& \frac 1 2 \left(\inf_{s\in[0,1]}\exp\left\{-\frac {N_{\rm S}(N_{\rm S}+1)} {2(N_{\rm B}'+1)^2} \frac{\sum_j \omega_j\left(\sqrt{\eta_j^{(0) }}-\sqrt{\eta_j^{(1) }}\right)^2}{\Lambda_{s}(2N_{\rm S}+1)+\Lambda_{1-s}(2N_{\rm S}+1)}|\overline{\bs x}|^2\right\}\right)^\nu+ \mathcal O\left(\frac{ m^2\nu (\overline{\bs x_q}+\overline{\bs x_p})^2 }{m_{\rm re}^2N_{\rm B}^3}\right)\\
\le & \frac 1 2 \exp\left\{-\frac {N_{\rm S}(N_{\rm S}+1)} {4(N_{\rm B}'+1)^2} \frac{\sum_j \omega_j\left(\sqrt{\eta_j^{(0) }}-\sqrt{\eta_j^{(1) }}\right)^2}{\Lambda_{\frac 1 2 }(2N_{\rm S}+1)}\sum_n |{\bs x}_n|^2\right\}+ \mathcal O\left(\frac{ m^2\nu(\overline{\bs x_q}+\overline{\bs x_p})^2  }{m_{\rm re}^2N_{\rm B}^3}\right)
\end{align}

Given the probability distribution 
\begin{align}
p(\{\bs x_n\})\approx \frac 1 {[4 (N_{\rm B}'+1)\pi ]^\nu}\exp\left[-\frac{\sum_n|\bs x _n|^2}{4(N_{\rm B}'+1)}\right]
\end{align}
for a constant $N_{\rm B}'$ due to the assumption $N_{\rm S}\ll N_{\rm B}$, we have the average error is: 
\begin{align}
p_{\rm hp} \left(\rho_{\bs \eta^{(0)}}^{\otimes \nu},\rho_{\bs \eta^{(1)}}^{\otimes \nu}\right)
\le & \frac 1 2 \left[1+ \frac{N_{\rm S}(N_{\rm S}+1)\sum_j \omega_j\left(\sqrt{\eta_j^{(0) }}-\sqrt{\eta_j^{(1) }}\right)^2}{(N_{\rm B}'+1)\Lambda_{\frac 1 2 }(2N_{\rm S}+1)}\right]^{-\nu}+ \mathcal O\left(\frac{m^2\nu  }{m_{\rm re}^2N_{\rm B}^2}\right)\\
= & \frac 1 2 \exp\left[- \frac{\nu N_{\rm S}(N_{\rm S}+1)\sum_j \omega_j\left(\sqrt{\eta_j^{(0) }}-\sqrt{\eta_j^{(1) }}\right)^2}{(N_{\rm B}'+1)\Lambda_{\frac 1 2 }(2N_{\rm S}+1)}\right]+ \mathcal O\left(\frac{m^2\nu}{m_{\rm re}^2N_{\rm B}^2}\right)
\end{align}
which has the same scaling with the single parameter case \cite{shi2022fulfilling}.

\subsection{Benchmark for pattern classification}\label{APP:PCC}

Classical pattern classification protocols can be modeled as the situation where the input source for the QI network will be described by a state with positive P-representation \cite{weedbrook2012gaussian,shapiro2020quantum}. Equivalently, the input source can be modeled by a statistical mixture of multi-mode coherent states. Given the result in Ref. \cite{zhuang2020entanglement}, we know that the minimum error is achieved by pure probe states in the quantum discrimination of channels. Therefore, with $2m$ coherent probe states $\otimes_{j=0}^{2m-1}|\beta_j\>$, the output state of the QI network will be a single-mode displaced thermal state $\rho_{\bs \beta}=D\left(\sum_{k=0}^{m-1}\sqrt{\omega_{k}\eta_{k}}e^{-i\theta_{k}}\beta_{k}\right)\rho_{N_{\rm B}}D\left(\sum_{l=0}^{m-1}\sqrt{\omega_{l}\eta_{l}}e^{-i\theta_{l}} \beta_{l}\right)$, which induces the relations: 
\begin{align}
\Tr\left[\rho_{\bs \eta_0,{\rm c}}^s\rho_{\bs \eta_1,{\rm c}}^{1-s}\right]&= \frac{2  G_{ s} (2N_{\rm B}+1)G_{1-s} (2N_{\rm B}+1)}{\Lambda_{ s}(2N_{\rm B}+1)+\Lambda_{1-s}(2N_{\rm B}+1)} \exp\left\{-\frac {2\left|\sum_{j}\sqrt{\omega_{j}}\left(\sqrt{\eta_{ j}^{(0)}}-\sqrt{\eta_{ j}^{(1)}}\right)\beta_j\right|^2} {\Lambda_{ s}(2N_{\rm B}+1)+\Lambda_{1-s}(2N_{\rm B}+1)} \right\}, \label{e1}
\end{align}
where $G_p(\mu)$ and $\Lambda_p(\mu)$ are functions defined as follows: 
\begin{align}
G_p(\mu)&=\frac{2^p}{(\mu+1)^p-(\mu-1)^p},\\
\Lambda_p(\mu)&=\frac{(\mu+1)^p+(\mu-1)^p}{(\mu+1)^p-(\mu-1)^p} .
\end{align}

Furthermore, it can be proved that Eq. (\ref{e1}) achieves its minimal value by with $s=\frac 1 2$ \cite{shi2022fulfilling}. Therefore, we have: 
\begin{align}
p_{\rm hp} \left(\rho_{\bs \eta^{(0)},{\rm c}}^{\otimes \nu},\rho_{\bs \eta^{(1)},{\rm c}}^{\otimes \nu}\right)&=\frac 1 2\min_{\bs \beta}  \exp\left\{-\frac {\nu\left|\sum_{j}\sqrt{\omega_{j}}\left(\sqrt{\eta_{ j}^{(0) }}-\sqrt{\eta_{j}^{(1) }}\right)\beta_j\right|^2} {\Lambda_{\frac 1 2 }(2N_{\rm B}+1)} \right\}
\end{align}

In the case where either $\{\omega_j\}$ or $\{\eta_j\}$ are unknown, the experimenter can not concentrate the power of the probe modes to one that minimizes the error probability. Further, we have the following conditions: 
\begin{align}
\sum_{j=0}^{m-1}\sqrt{\omega_{j}}\left(\sqrt{\eta_{ j}^{(0) }}-\sqrt{\eta_{j}^{(1) }}\right)=\mathcal O(0),%\tag{S83}
\end{align}
we will have a scaling advantage using the CtoD quantum strategy.

\section{Discussion on average phase sensing}
\label{app:average_phase_dis}

Let's reduce our discussion to single-parameter estimation. As a concrete example, we consider the sensing of the average phase: 
\begin{align}\label{eqf1}
\overline \theta = \sum_{j=0}^{m-1} \sqrt{\omega_j\eta_j}\,\theta_j.
\end{align}
Next, there will be two distinct scenarios: (i) The differences between each pair of the $m$ phases are known. (ii) There still exist $m-1$ independent and unknown  phases after estimation of the average phase. In the subsequent subsections, we shall examine these two scenarios separately. 

\subsection{Degenerative case}

If differences between each pair of phases are known, we may apply the following transformation \cite{suzuki2020quantum} of 
to the QFIM achievable by the QI network: 
\begin{align}
F^{\rm qi}_{\overline \theta}&=\left(\begin{matrix}
\frac{\partial \theta_0}{\partial \overline \theta },\cdots, \frac{\partial \theta_{m-1}}{\partial \overline \theta }
\end{matrix}\right)\overline F \left(\begin{matrix}
\frac{\partial \theta_0}{\partial \overline \theta }\\\vdots\\ \frac{\partial \theta_{m-1}}{\partial \overline \theta }
\end{matrix}\right)\\
&=\frac{4m\nu N_{\rm S}(N_{\rm S}+1)}{(N_{\rm B}'+1)(2N_{\rm S}+1)}+\mathcal O\left(\frac{m^2\nu N_{\rm S}^2}{m_{\rm re}N_{\rm B}^2}\right). 
\end{align}
where $\overline F$ is the QFIM obtained by the CtoD, as shown in Eq. (\ref{eq17}), $m_{\rm re}^{-1}$ is defined in Eq. (\ref{eq2}) in the main text and is given by the premise $\omega_j\sim m_{\rm re}^{-1}$. Then, its RMSE is: 
\begin{align}\label{eqf4}
\epsilon^{\rm qi}_{\overline \theta}=\sqrt{\frac{(N_{\rm B}'+1)(2N_{\rm S}+1)}{4m\nu N_{\rm S}(N_{\rm S}+1)}}+\mathcal O\left(\sqrt{\frac{m N_{\rm S}}{m_{\rm re}^2\nu N_{\rm B}}}\right).
\end{align}

On the other hand, given by the convexity of Fisher information \cite{liu2020quantum}, 
the output state from classical illumination network in Eq. (\ref{eqa4}) will achieve the maximal value of quantum Fisher information (QFI) when the probe is pure. By using the results of QFI for Gaussian states \cite{gao2014bounds,nichols2018multiparameter,vsafranek2018estimation,liu2020quantum}, we will have the maximal QFI for classical case: 
\begin{align}
\map F^{\rm ci}_{\overline \theta}&=\max_{\{\beta_j;|\beta_j|=N_{\rm S}\}}\frac{4}{2N_{\rm B}+1}\left|\frac{\partial \sum_{j=0}^{m-1}\sqrt{\omega_j\eta_j}e^{-i\theta_j}\beta_j}{\partial \overline \theta }\right|^2,\\
&=\frac{4m^2 N_{\rm S}}{2N_{\rm B}+1}.
\end{align}
The corresponding root mean-square-error (RMSE) achievable by the classical  network is: 
\begin{align}\label{eqf7}
\epsilon_{\overline \theta}^{\rm ci}&=\sqrt{\frac{2N_{\rm B}+1}{4m^2 \nu N_{\rm S}}}. 
\end{align}

As shown in Eqs. (\ref{eqf4}) and (\ref{eqf7}), there is a disadvantage of QI network average parameter sensing in the scaling of transmitter number $m$. 

\subsection{Non-degenerative case}

Consider the case where there are $m$ unknown independent parameters $\{\overline \theta,\theta_1,\cdots,\theta_{m-1}\}$, where $\overline \theta$ is defined in Eq. (\ref{eqf1}). The partial Fisher information of $\overline \theta$ can be obtain by computing the Schur's complement \cite{suzuki2020quantum} of the initial FIM, i.e. that in Eqs. (\ref{eq17}). The presence of $m-1$ nuisance parameters in both quantum and classical scenarios leads to a reduction in the value of partial Fisher information. Given the current focus on multi-parameter sensing, we will leave this research question for future works.

\end{widetext}

\end{document}